\documentclass[reprint,superscriptaddress,amsmath,amssymb,aps,prx]{revtex4-2}
\usepackage{graphicx}
\usepackage{dcolumn}
\usepackage{bm}
\usepackage{qcircuit}
\usepackage{algpseudocode}
\usepackage{algorithm}
\usepackage{subfigure}
\usepackage[pass]{geometry}
\usepackage{float}
\usepackage{enumitem}
\usepackage{bibunits}
\defaultbibliography{ref.bib}
\defaultbibliographystyle{unsrt}
\makeatletter
\renewcommand*\env@matrix[1][\arraystretch]{%
  \edef\arraystretch{#1}%
  \hskip -\arraycolsep
  \let\@ifnextchar\new@ifnextchar
  \array{*\c@MaxMatrixCols c}}
\makeatother
\begin{document}
\preprint{APS/123-QED}
\title{A polynomial-time dissipation-based quantum algorithm for solving the ground states of a class of classically hard Hamiltonians}
\author{Zhong-Xia Shang}
\email{ustcszx@mail.ustc.edu.cn}
\affiliation{Hefei National Research Center for Physical Sciences at the Microscale and School of Physical Sciences,
University of Science and Technology of China, Hefei 230026, China}
\affiliation{Shanghai Research Center for Quantum Science and CAS Center for Excellence in Quantum Information and Quantum Physics,
University of Science and Technology of China, Shanghai 201315, China}
\affiliation{Hefei National Laboratory, University of Science and Technology of China, Hefei 230088, China}
\author{Zi-Han Chen}
\affiliation{Hefei National Research Center for Physical Sciences at the Microscale and School of Physical Sciences,
University of Science and Technology of China, Hefei 230026, China}
\affiliation{Shanghai Research Center for Quantum Science and CAS Center for Excellence in Quantum Information and Quantum Physics,
University of Science and Technology of China, Shanghai 201315, China}
\affiliation{Hefei National Laboratory, University of Science and Technology of China, Hefei 230088, China}
\author{Chao-Yang Lu}
\affiliation{Hefei National Research Center for Physical Sciences at the Microscale and School of Physical Sciences,
University of Science and Technology of China, Hefei 230026, China}
\affiliation{Shanghai Research Center for Quantum Science and CAS Center for Excellence in Quantum Information and Quantum Physics,
University of Science and Technology of China, Shanghai 201315, China}
\affiliation{Hefei National Laboratory, University of Science and Technology of China, Hefei 230088, China}
\author{Jian-Wei Pan}
\affiliation{Hefei National Research Center for Physical Sciences at the Microscale and School of Physical Sciences,
University of Science and Technology of China, Hefei 230026, China}
\affiliation{Shanghai Research Center for Quantum Science and CAS Center for Excellence in Quantum Information and Quantum Physics,
University of Science and Technology of China, Shanghai 201315, China}
\affiliation{Hefei National Laboratory, University of Science and Technology of China, Hefei 230088, China}
\author{Ming-Cheng Chen}
\email{cmc@ustc.edu.cn}
\affiliation{Hefei National Research Center for Physical Sciences at the Microscale and School of Physical Sciences,
University of Science and Technology of China, Hefei 230026, China}
\affiliation{Shanghai Research Center for Quantum Science and CAS Center for Excellence in Quantum Information and Quantum Physics,
University of Science and Technology of China, Shanghai 201315, China}
\affiliation{Hefei National Laboratory, University of Science and Technology of China, Hefei 230088, China}

\begin{abstract}
In this work, we give a polynomial-time quantum algorithm for solving the ground states of a class of classically hard Hamiltonians. The mechanism of the exponential speedup that appeared in our algorithm comes from dissipation in open quantum systems. To utilize the dissipation, we introduce a new idea of treating vectorized density matrices as pure states, which we call the vectorization picture. By doing so, the Lindblad master equation (LME) becomes a Schrödinger equation with non-Hermitian Hamiltonian. The steady state of the LME, therefore, corresponds to the ground states of a special class of Hamiltonians. The runtime of the LME has no dependence on the overlap between the initial state and the ground state. For the input part, given a Hamiltonian, under plausible assumptions, we give a polynomial-time classical procedure to judge and solve whether there exists LME with the desired steady state. For the output part, we propose a novel measurement strategy to extract information about the ground state from the original steady density matrix. We show that the Hamiltonians that can be efficiently solved by our algorithms contain classically hard instances assuming $\text{P}\neq \text{BQP}$. We also discuss possible exponential complexity separations between our algorithm and previous quantum algorithms without using the vectorization picture.

\end{abstract}
\maketitle
\section{Introduction}
Quantum computers can utilize properties of quantum mechanics like superposition and entanglement to solve problems faster than their classical counterparts. Thus, over the past decades, there have been vast studies on quantum algorithms and many important ones have been found \cite{montanaro2016quantum}. Among the studies, quantum algorithms on solving quantum ground states are of particular interest due to their potential vast applications in many-body systems \cite{thouless2014quantum}, quantum chemistry \cite{levine2009quantum}, classical combinatorial optimization problems \cite{ausiello2012complexity} and machine learning \cite{jordan2015machine}. However, it has been proved that general $k$-local Hamiltonian ground energy problems belong to QMA complexity class i.e. the quantum version of NP \cite{kempe2006complexity}, which indicates that ground state problems are even difficult for quantum computers.

In this work, we mainly focus on the task of estimating expectation values of arbitrary operators with respect to the ground state, which generally requires quantum algorithms for ground-state preparation rather than estimating ground energy. Note that these two tasks can have different complexities since estimating ground energy doesn't necessarily require ground state preparations \cite{chen2024sparse}. We now give a brief review of existing quantum algorithms on this task. A large class of ground-state preparation algorithms all share similar ideas of projecting out the ground state from an initial state. A standard method is to use quantum phase estimation \cite{kitaev1995quantum} combined with amplitude amplification \cite{brassard2002quantum}. Given a Hamiltonian with known ground energy whose spectral gap between the ground state and the first excited state is bounded by $\Delta$, the required runtime to prepare its ground state to a fidelity $1-\varepsilon$ is of order $\mathcal{O}(\Delta^{-1}\varepsilon^{-1}\zeta^{-2})$ with $\zeta$ the overlap between the initial state and the ground state. This runtime, however, is far from optimal and can be dramatically improved by techniques such as linear combinations of unitaries \cite{childs2012hamiltonian} and quantum signal processing \cite{low2017optimal} (and its variants \cite{gilyen2019quantum,dong2022ground}). By using these techniques, various works \cite{poulin2009preparing,ge2019faster,lin2020near,dong2022ground,zeng2109universal} have reduced the complexity of solving this task to $\mathcal{O}(\Delta^{-1}\log(\varepsilon^{-1})\zeta^{-1})$.

There are also several types of quantum algorithms for the ground-state preparation task without rigorous proof on the complexity. A prominent one is the adiabatic quantum computing \cite{farhi2000quantum} whose rigorous complexity for ground state problems is still an open question \cite{van2001powerful} since it is very difficult to evaluate the gap during the adiabatic path. There is also a popular type called the hybrid classical-quantum algorithms including the quantum approximate optimization algorithm \cite{farhi2014quantum}, variational quantum eigensolver \cite{peruzzo2014variational} which are promising candidates in the NISQ era \cite{preskill2018quantum} due to their low requirements on quantum circuits. The complexity of this type is unclear due to the classical optimization procedure. Indeed, these algorithms involve classically minimizing highly non-convex cost functions with barren plateaus \cite{mcclean2018barren} which has been shown to be NP-hard \cite{bittel2021training}. Another type is the quantum imaginary time evolution \cite{motta2020determining} algorithm which tries to find unitary circuits to approximate imaginary time evolution. However, to make the classical procedure efficient, it only works for those with bounded correlation length during the imaginary time evolution. Thus, it is difficult to give complexity dependence on the final fidelity. 

From the $\mathcal{O}(poly(\zeta^{-1}))$ scaling of the runtime in the above algorithms, we can see why solving ground state can be even inefficient on quantum computers \cite{bennett1997strengths,zalka1999grover}. As the number of qubits $n$ of the system grows, with no prior knowledge, the overlap between an initial state and the ground state is exponentially small, thus the runtime grows exponentially with the qubit number. Recently, a new type of algorithm \cite{cubitt2023dissipative,ding2023single} utilizing dissipation in open quantum systems for ground-state preparation has been proposed. These algorithms construct carefully designed dissipation such that the fixed point (steady state) of the system is the ground state of the Hamiltonian. By doing so, the runtime dependence on the initial overlap $\zeta$ can be removed, which, however, requires a new dependence on the mixing time as the price. Analyzing the mixing time is a rather difficult task, and thus, no clear results on whether these algorithms have advantages over previous ones. Nevertheless, utilizing dissipation for ground-state preparation is rather attractive since it is possible to use the exponential behavior of decoherence to counter the exponential scaling of the dimension of Hilbert space.

In this work, we give a new quantum algorithm by utilizing dissipation to solve the ground states of a restricted class of Hamiltonians. While general Hamiltonians are difficult for quantum computers, we find there exists Hamiltonians that can be efficiently solved by our algorithm and unlikely can be efficiently solved classically. The main task we set is estimating the expectation values of observables with respect to the ground state. The special feature of our algorithm is that it can extract ground state information without truly preparing the ground state by introducing a vectorization picture of treating the vectorized density matrices as pure states. When considering other tasks such as amplitude estimation, we will show there also exist exponential complexity separations between our algorithm and previous quantum algorithms without using the vectorization picture. Later, we will discuss several aspects of the algorithm including the symmetry of solvable Hamiltonians, generalizing to other types of Hamiltonians, and the "non-linear`` dynamics in the algorithm. 

\section{Algorithm}
\subsection{Lindblad master equation and Hamiltonian correspondence}
We begin to introduce our algorithm. The first component comes from the dynamics of open quantum systems. Consider putting a system in an environment which is large enough such that the Markovian approximation is valid, then the dynamics of the system is governed by the Lindblad master equation (LME) \cite{gardiner2004quantum,haroche2006exploring}:
\begin{equation}\label{1}
\frac{d\rho}{dt}=\mathcal{L}[\rho]=-i[H,\rho]+
\sum_i \lambda_i(F_i\rho F_i^\dag-
\frac{1}{2}\{\rho,F_i^\dag F_i\})
\end{equation}
where $\rho=\sum_{ij}\rho_{ij}|i\rangle\langle j|$ is the density matrix of the system and $F_i$ are quantum jump channels with strength $\lambda_i$. LME describes the dissipative nature of a system coupled with an environment. To see this more clearly, we can re-express the above LME as a vector form \cite{albert2014symmetries,minganti2018spectral}:
\begin{equation}\label{2}
\quad\frac{d\vec{\rho}}{dt}=L\vec{\rho}
\end{equation}
where $\vec{\rho}=\sum_{ij}\rho_{ij}|i\rangle|j\rangle$ is the vector representation of the density matrix $\rho$ and $L$ is the Liouvillian generator for the LME semi-group which are not Hermitian in general has the following matrix form:
\begin{eqnarray}\label{3}
&&L=(-i(H\otimes I-I\otimes H^T)+\sum_i \lambda_i D[F_i])\\
&&\text{where}\quad D[F_i]=F_i\otimes F_i^*-\frac{1}{2}F_i^\dag F_i\otimes I-I\otimes\frac{1}{2}F_i^TF_i^*\nonumber
\end{eqnarray}
Eq. \ref{2} is attractive as it can be understood as a Schrödinger equation with non-Hermitian Hamiltonian $iL$ which thus contains a natural imaginary time evolution. If the LME has a steady density matrix $\rho_{ss}$ i.e. $\mathcal{L}[\rho_{ss}]=0$, then the corresponding $\vec{\rho_{ss}}$ is the unnormalized ground state of Hermitian Hamiltonian $L^\dag L$ with zero ground energy. Thus, the information of the ground state of $L^\dag L$ is contained in $\rho_{ss}$. It can be shown that the ground state subspace of $L^\dag L$ is identical to the (vectorized) steady state subspace of $L$ even for degenerate cases \cite{minganti2018spectral}.

This part is the main resource for the potential exponential speedup. Naively, when $L$ is Hermitian, LME corresponds to a natural imaginary time evolution such that the runtime scaling with respect to the overlap $\zeta$ between the initial state $\rho_0$ and the steady state $\rho_{ss}$, the gap $\Delta_l$ of $L$ denoting the smallest real part of the gaps between the steady state and other eigenvectors of $L$ (assuming $L$ is diagonalizable \cite{minganti2018spectral}), and the required final overlap $1-\varepsilon$ to be achieved is $\mathcal{O}(\Delta_l^{-1}log(\varepsilon^{-1/2}\zeta^{-1}))$. Here, the overlap denotes the overlap between vectorized and normalized density matrices defined in the following section. Since for the Hermitian case, $\zeta$ has a non-zero lower bound $2^{-n/2}$, thus the dependence over zeta can be removed to give an upper bound on the required time: $t\geq\Delta_l^{-1}(\frac{\ln (2)}{2}n+\ln(\varepsilon^{-\frac{1}{2}}))$. For non-Hermitian cases, we typically use the mixing time $t_{mix}$ to replace $\Delta_l$ and $\zeta$ to obtain a similar bound: $t\geq \frac{t_{mix}(n+log(\varepsilon^{-1}))}{2}$. The mixing time $t_{mix}$ is defined as the smallest time such that:
\begin{eqnarray}\label{mix}
\|e^{Lt_{mix}} (\rho_1-\rho_2)\|_1\leq \frac{1}{2}\|\rho_1-\rho_2\|_1\text{, for any $\rho_1$ and $\rho_2$}
\end{eqnarray}
The reason to replace the spectral gap by the mixing time is that while a polynomial large $t_{mix}$ necessarily indicates a polynomial small $\Delta_l$, the inverse claim is not always true \cite{temme2010chi} and only holds for special cases \cite{chen2023quantum,verstraete2009quantum}.

There have been various proposals \cite{kliesch2011dissipative,childs2016efficient,cleve2016efficient,li2022simulating,patel2023wave} for simulating LME by quantum circuits. Typically, starting from an initial state $|0\rangle_a|s_0\rangle$, these proposals implement carefully designed $U_t$ such that $Tr_a(U_t|0\rangle_a|s_0\rangle\langle 0|_a\langle s_0|U_t^\dag)=e^{\mathcal{L}t}[|s_0\rangle\langle s_0|]$. Suppose the goal is to simulate LME dynamics for a time $t$ with a precision $\delta$, current state-of-the-art algorithms \cite{cleve2016efficient,li2022simulating} have achieved a complexity of order $\mathcal{O}(t\text{polylog}(t/\delta))$. Thus, LME can be efficiently simulated on quantum computers and throughout this work, we will only focus on the bare runtime of LME and ignore its circuit implementations.

\subsection{Output: Evaluating expectation values}
Since the information of the ground state of $L^\dag L$ is contained in $\rho_{ss}$, if we merely want to obtain expectation values of some operator $A$ with respect to $|\rho_{ss}\rangle$: $\langle A\rangle=\langle \rho_{ss}| A|\rho_{ss}\rangle$, we now show that this can be done by measurements on $\rho_{ss}$ with no need to actually prepare the ground state. To do so, we introduce the following mapping:
\begin{equation}\label{4}
f:\quad\rho\rightarrow |\rho\rangle
\end{equation} 
where $|\rho\rangle$ is defined as $\frac{1}{C_{\rho}}\sum_{ij}\rho_{ij}|i\rangle|j\rangle$ with the normalization factor $C_\rho=||\rho||_F=\sqrt{\sum_{ij}|\rho_{ij}|^2}$. The key is how to understand this mapping. Here, this mapping means that we are treating density matrices as pure states. For example, a single-qubit maximum mixed density matrix $(|0\rangle\langle 0|+|1\rangle\langle 1|)/2$ is treated as a two-qubit Bell state $(|0\rangle|0\rangle+|1\rangle|1\rangle)/\sqrt{2}$. We will call the subsystem labeled by index $i$ the row subsystem and the subsystem labeled by index $j$ the column subsystem. This mapping connects $\rho$ and $|\rho\rangle$ by the following relation:
\begin{eqnarray}\label{5}
\langle\rho|A|\rho\rangle=\frac{Tr(B\rho\otimes\rho)}{Tr(\rho^2)}
\end{eqnarray}
where each matrix element of $B$ $B_{il,jk}=\langle i|\langle l|B|j\rangle |k\rangle$ has the following relation with $A$:
\begin{equation}\label{6}
B_{il,jk}=A_{ij,kl}
\end{equation}
Eq. \ref{5} means the information of $|\rho\rangle$ i.e. the value of $\langle\rho|A|\rho\rangle$ with $A$ a Hermitian operator can be obtained from $\rho$ by the value of the ratio of an operator $B$'s expectation value under $\rho\otimes\rho$ to the purity of $\rho$. We name $B$ as the substitute operator of $A$. 

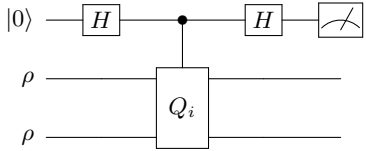
\begin{figure}[htb]
\centering
\[ \Qcircuit @C=1.5em @R=1em @!R{
        \lstick{|0\rangle} & \gate{H} & \ctrl{1} & \gate{H} & \meter\\
        \lstick{\rho} & \qw & \multigate{1}{Q_i} & \qw & \qw\\
        \lstick{\rho} & \qw & \ghost{Q_i} & \qw & \qw
        }\]
\caption{The Hadamard test circuit for $Re(Tr(Q_i\rho\otimes\rho))$.}
\label{fig3}
\end{figure}

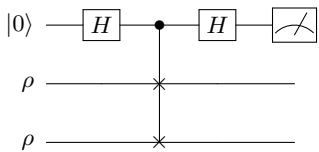
\begin{figure}[htb]
    \centering
    \[ \Qcircuit @C=1.5em @R=1em @!R{
            \lstick{|0\rangle} & \gate{H} & \ctrl{1} & \gate{H} & \meter\\
            \lstick{\rho} & \qw & \qswap & \qw & \qw\\
            \lstick{\rho} & \qw & \qswap\qwx & \qw & \qw
            }\]      
\caption{The Swap test circuit for $Tr(\rho^2)$.}   
\label{fig4}
\end{figure}

Having the formula Eq. \ref{5}, the following questions are how to measure its right-hand side and how efficient the measurement can be. Before introducing the measurement procedure, a prior thing to show is that the tensor product properties of $A$ are not lost in $B$. More clearly, if each index of $A$ in Eq. \ref{6} actually contains indexes of $n$ qubits e.g. $i\rightarrow i_1i_2...i_n$, then the following relation is satisfied:
\begin{eqnarray}\label{7}
&&\text{if:}\quad A_{i_1i_2...i_nj_1j_2...j_n,k_1k_2...k_nl_1l_2...l_n}=\nonumber\\&&\quad A^1_{i_1j_1,k_1l_1}A^2_{i_2j_2,k_2l_2}...A^n_{i_nj_n,k_nl_n},\nonumber\\
&&\text{then:} \quad B_{i_1i_2...i_nl_1l_2...l_n,j_1j_2...j_nk_1k_2...k_n}=\nonumber\\&&\quad B^1_{i_1l_1,j_1k_1}B^2_{i_2l_2,j_2k_2}...B^n_{i_nl_n,j_nk_n} 
\end{eqnarray}
where relations between $A^1,A^2,...A^n$ and $B^1,B^2,...B^n$ satisfy the rules in Eq. \ref{6}. Due to the relation in Eq. \ref{6}, $n>1$ situations can be generalized from the basic $n=1$ case where $A$ is a 2-qubit operator. For this case, the Hermitian $A$ can be expressed as a real linear combination of 16 2-qubit Pauli operators. Each operator has a corresponding $B$ operator which we will call the 2-qubit Pauli substitute operator. Interestingly, although the Hermiticity of 2-qubit Pauli operators is lost in the 2-qubit Pauli substitute operators, the unitarity is not, i.e. each 2-qubit Pauli substitute operator is unitary. All 16 2-qubit Pauli substitute operators are summarized in Table. \ref{tab1}.
\begin{table}[t]
  \centering
  \begin{ruledtabular}
  \begin{tabular}{c|ccc}
    ID&$A$& $B$ & Spectra of $B$ \\\hline
    1&$II$& $0.5 II+0.5XX+0.5YY+0.5ZZ$& \{1,1,1,-1\}\\
    2&$XX$& $0.5 II+0.5XX-0.5YY-0.5ZZ$& \{1,1,-1,1\}\\
    3&$YY$& $-0.5 II+0.5XX-0.5YY+0.5ZZ$& \{1,-1,-1,-1\}\\
    4&$ZZ$& $0.5 II-0.5XX-0.5YY+0.5ZZ$& \{1,-1,1,1\}\\\hline
    5&$IX$& $0.5 IX+0.5XI+0.5iYZ-0.5iZY$& \{-1,$i$,-$i$,1\}\\
    6&$XI$& $0.5 IX+0.5XI-0.5iYZ+0.5iZY$& \{-1,$i$,-$i$,1\}\\
    7&$YZ$& $-0.5i IX+0.5iXI+0.5YZ+0.5ZY$& \{-1,$i$,-$i$,1\}\\
    8&$ZY$& $-0.5i IX+0.5iXI-0.5YZ-0.5ZY$& \{-1,$i$,-$i$,1\}\\\hline
    9&$IY$& $-0.5 IY+0.5iXZ-0.5YI-0.5iZX$& \{-1,$i$,-$i$,1\}\\
    10&$YI$& $0.5 IY+0.5iXZ+0.5YI-0.5iZX$& \{-1,$i$,-$i$,1\}\\
    11&$XZ$& $0.5i IY+0.5XZ-0.5iYI+0.5ZX$& \{-1,$i$,-$i$,1\}\\ 
    12&$ZX$& $-0.5i IY+0.5XZ+0.5iYI+0.5ZX$& \{-1,$i$,-$i$,1\}\\\hline 
    13&$IZ$& $0.5 IZ+0.5iXY-0.5iYX+0.5ZI$& \{$i$,-$i$,1,-1\}\\
    14&$ZI$& $0.5 IZ-0.5iXY+0.5iYX+0.5ZI$& \{$i$,-$i$,1,-1\}\\   
    15&$XY$& $0.5i IZ-0.5XY-0.5YX-0.5iZI$& \{1,-1,-$i$,$i$\}\\
    16&$YX$& $0.5i IZ+0.5XY+0.5YX-0.5iZI$& \{1,-1,-$i$,$i$\}\\
  \end{tabular}
  \end{ruledtabular}
  \caption{The 16 2-qubit Pauli operators ($A$) and their corresponding 2-qubit Pauli substitute operators ($B$). Each $B$ is unitary whose eigenvalues are presented.}
  \label{tab1}
\end{table}
The unitarity of 2-qubit Pauli substitute operators makes sure that we can measure the expectation $Tr(B\rho\otimes\rho)$ by Hadamard tests \cite{aharonov2009polynomial}. 
Concretely, consider a to be measured 2$n$-qubit Hermitian operator expressed as a combination of $m$ terms:
\begin{equation}\label{8}
    A=\sum_{i=1}^m g_iP_i
\end{equation}
where $P_i$ are $2n$-qubit Pauli operators and $g_i$ are the strengths which are real numbers. The substitute operator of $A$ in Eq. \ref{8} can be expressed as a similar form: 
\begin{equation}\label{9}
    B=\sum_{i=1}^m g_iQ_i
\end{equation}
where due to the transformation rule Eq. \ref{6}, the tensor product relation and the unitarity of the 2-qubit Pauli substitute operators, each $2n$-qubit Pauli substitute operator $Q_i$ is unitary. Thus, the Hadamard test of $\mathcal{O}(1)$ depth can be used to evaluate each $Re(Tr(Q_i\rho\otimes\rho))$ (Fig. \ref{fig3}). 
The $\sigma_z$ expectation value of the ancillary qubit gives the value of $Re(Tr(Q_i\rho\otimes\rho))$. By multiplying each estimated value of $Re(Tr(Q_i\rho\otimes\rho))$ by its weight $g_i$ and summing up the results, we can obtain an estimation of the numerator $Tr(B\rho\otimes\rho)$ in Eq. \ref{5}.  There is no need to measure the imaginary value of each $Q_i$ since they will cancel out eventually guaranteed by the Hermiticity of $A$. 
For the purity $Tr(\rho^2)$ in the denominator, Swap tests \cite{buhrman2001quantum}, shown in Fig. \ref{fig4}, of $\mathcal{O}(1)$ depth are used to obtain the purity $Tr(\rho^2)$ in the denominator of Eq. \ref{5} as $2p_s-1$ where $p_s$ is the probability of measuring the ancillary qubit in the $|0\rangle$ state.
Thus, by using Hadamard tests and Swap tests, we can measure the value of $\langle \rho|A|\rho\rangle$. We can show that the sampling complexity is $\mathcal{O}(m\gamma^{-2}\epsilon^{-2})$ to achieve a mean squared error containing both bias and variance \cite{james2013introduction} within $\epsilon$ assuming the purity $Tr(\rho_{ss}^2)= \gamma$. Here, we only think about the standard quantum limit estimation for simplicity, and one can always convert it into Heisenberg limit estimations \cite{rall2020quantum,knill2007optimal}.

\begin{figure*}[htbp]
\centering
\includegraphics[width=0.7\textwidth]{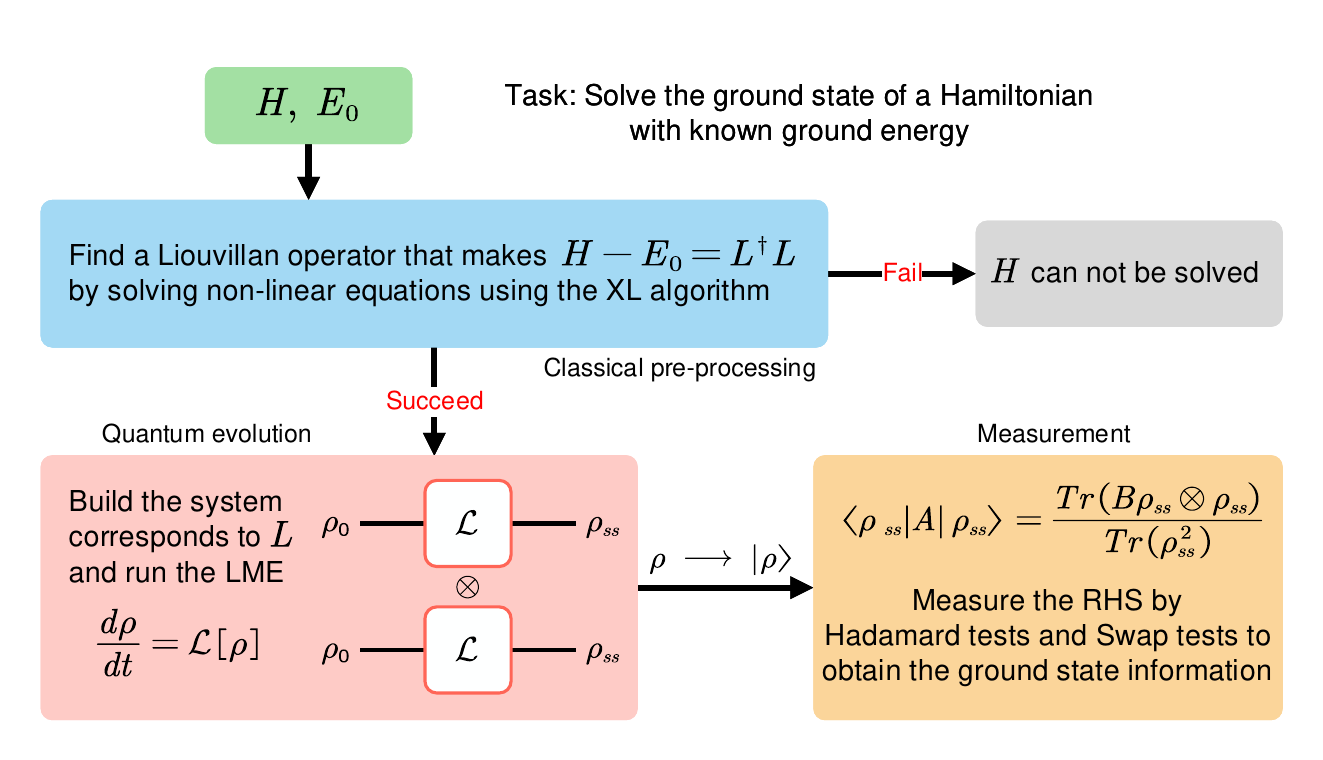}
\caption{The structure of the algorithm. Given a Hamiltonian $H$ with known ground energy $E_0$, the judgment on whether there exists $L$ makes $H-E_0=L^\dag L$ is done by the XL algorithm. If the XL algorithm gives a solution $L$, then we can generate a quantum system whose dynamics are governed by an LME whose corresponding Liouvillian generator is $L\otimes L$. Then, let the system evolve freely to the steady state $\rho_{ss}\otimes\rho_{ss}$ of the LME (if the steady state exists). Next, if the task is to obtain expectation values of some operators with respect to the ground state $|\rho_ss\rangle$, we can do measurements on $\rho_{ss}$ for the Hadamard test and the Swap test based on Eq. \ref{5}.}
\label{fig1}
\end{figure*}

\subsection{Input}
Currently, the Hamiltonian $L^\dag L$ we solved is calculated from the given $L$, which is easy as long as there are only polynomial terms in $L$. However, given a Hamiltonian $H$ with given ground energy $E_0$ first, the reverse procedure of getting $L$ that makes $H-E_0=L^\dag L$ is non-trivial. This is NP-hard in general since it will require solving exponentially large systems of multivariate quadratic equations \cite{garey1979computers}. Fortunately, if we add some restrictions on $L$ and $L^\dag L$ such as locality (in terms of Pauli weight) or connectivity constraints which are very practical on real quantum computers, we can get the corresponding $L$ in only polynomial-time, which is the final component of our algorithm. The reason for this is that in these cases, we only need to consider a polynomial number of Pauli terms and we only need to solve a polynomially large system of over-defined multivariate quadratic equations which can be empirically efficiently solved by a classical algorithm called the XL algorithm \cite{courtois2000efficient,bardet2005asymptotic}. The XL algorithm states that if the number of the unknowns $N_u$ and the number of the equations $N_e$ satisfy $N_e\geq N_u^2 r$ for all $0<r\leq 1/2$, the runtime of the XL algorithm is about $N_u^{\mathcal{O}(1/\sqrt{r})}$. Specifically, we proved that the runtime of the XL algorithm is around $N_u^{5.678}$ for 4-local $H$, $N_u^{8.111}$ for 6-local $H$, etc. Here, we call an operator to be $k$-local if every Pauli term of it has {$X$, $Y$, $Z$} act on at most $k$ qubits.

\subsection{Workflow}
Having introduced the three components of our algorithm, we can now formally describe our algorithm as shown in Fig. \ref{fig1}. Given a Hamiltonian $H$ with known ground energy $E_0$, we first judge and solve whether there exists $L$ makes $H-E_0=L^\dag L$. If the XL algorithm gives a solution $L$, then we can generate a quantum system simulating LME dynamics governed by $L\otimes L$. Then, let the system evolve to the steady state $\rho_{ss}\otimes\rho_{ss}$ of the LME. Next, we can do the measurement procedure based on Eq. \ref{5} to estimate the expectation values of some operators with respect to $|\rho_{ss}\rangle$. 

\section{Complexity discussion}

\subsection{Quantum easiness}
The basic belief for quantum easiness is the possibility to utilize the exponentially fast dissipative dynamics to counter the exponentially large Hilbert space (or exponentially small initial overlap). The total runtime of our algorithm depends on the runtime of the XL algorithm, the runtime of LME evolution, and the runtime of the output part. From the above discussions, we can conclude that to make sure the runtime of our algorithm is within the polynomial regime, we require:
\begin{itemize}[leftmargin=*,itemsep=2pt,topsep=0pt,parsep=0pt]
\item[\romannumeral1] Input requirement: $H$ is local or has other restrictions to satisfy the over-defined regime for the XL algorithm.
\item[\romannumeral2] LME requirement: The mixing time $t_{mix}$ is of order $\mathcal{O}(poly(n))$ (which in many cases is equivalent to the gap $\Delta_l$ of $L$ is of order $\mathcal{O}(poly(n)^{-1})$). 
\item[\romannumeral3] Output requirement: The purity of the steady state $\gamma=Tr(\rho_{ss}^2)$ is of order $\mathcal{O}(poly(n)^{-1})$.
\end{itemize}
we want to emphasize here that the gap $\Delta_l$ generally has no direct relation with the gap of $L^\dag L$. Since the gap of $L^\dag L$ is related to the gap of $L$ in terms of singular value decomposition while $\Delta_l$ is defined in terms of eigenvalue decomposition.

Compared with other dissipation-based quantum algorithms \cite{cubitt2023dissipative,ding2023single} where dissipative dynamics are specially designed, it is easier to find quantum easiness in our algorithm since there are no restrictions on LME and each LME is encoded to a Hamiltonian. As a cost, our algorithm is not universal for arbitrary Hamiltonian.

\subsection{Classical hardness}
The basic belief for the classical hardness originated from the hardness of classically simulating LME, which is a plausible guarantee since either steady states of LME are easy to solve or our algorithm has advantages. We believe the latter is true which in some sense is similar to the motivation of quantum simulation \cite{feynman2018simulating}.

Interestingly, among Hamiltonians that satisfy all three conditions for quantum easiness, there are classically hard instances. A direct example can be found in Ref. \cite{verstraete2009quantum} where LME is designed to encode universal quantum circuits. Suppose we want to simulate an arbitrary quantum circuit: $|\psi_T\rangle=U_TU_{T-1}...U_1|0\rangle^{\otimes n}$ with all layers local and $T$ of order $\mathcal{O}(poly(n))$, we can then define an LME with no internal Hamiltonian and with two types of jump operators:
\begin{equation}\label{de10}
F_i=|0\rangle\langle 1|_i\otimes|0\rangle\langle 0|_a
\end{equation}
\begin{equation}\label{de11}
F_t=U_t\otimes|t+1\rangle\langle t|_a+U_t^\dag \otimes |t\rangle\langle t+1|_a
\end{equation}
with $i=1,...,n$ and $t=0,...,T$. This LME has a unique state state:
\begin{equation}\label{de12}
\rho_{ss}=\frac{1}{T+1}\sum_{t=0}^T|\psi_t\rangle\langle \psi_t|\otimes |t\rangle\langle t|_a
\end{equation}
In $\rho_{ss}$, each basis $|t\rangle_a$ of the ancilla time register is associated with the corresponding circuit output state $U_t...U_1|0\rangle^{\otimes n}$. Since estimating the probability $p_1$ of the first qubit in $|1\rangle$ under the state $|\psi_T\rangle$ up to additive precision is BQP-complete \cite{aharonov2017interactive,janzing2005ergodic}, and we can use:
\begin{equation}\label{de13}
p_1=\frac{1-(T+1)\langle \rho_{ss}|O|\rho_{ss}\rangle}{2}
\end{equation}
with $O=Z_1\otimes I_{n-1}\otimes |T\rangle\langle T|\otimes I_{n+a}$ to estimate $p_1$, which can be done efficiently by measurements based on Eq. \ref{5}, thus, assuming $\text{P}\neq \text{BQP}$, classical hardness is proved. It is easy to find this LME also satisfies all the conditions (mixing time of LME, purity of the steady state, and the locality of $L$) for our algorithm to be in a polynomial time. Thus, its corresponding $L^\dag L$ is exactly such an example for quantum-classical separation. Also, since now $L^\dag L$ encodes the output states of universal quantum computation into the ground state, it is another construction of circuit-to-Hamiltonian mapping \cite{kitaev2002classical,breuckmann2014space,anshu2024circuit}. For other instances, while there is no strong complexity-theoretic guarantee, we conjecture the average-case hardness of them since no efficient classical algorithms are known either.

\subsection{Comparison with quantum algorithms without vectorization}

Regarding solving the ground state $|\rho_{ss}\rangle$, previous quantum algorithms aim to prepare $|\rho_{ss}\rangle$ while our algorithm in the vectorization picture aims to prepare $\rho_{ss}$ but treat it as $|\rho_{ss}\rangle$. Since $|\rho_{ss}\rangle$ and $\rho_{ss}$ are physically different two quantum states, it is natural to ask whether our algorithm has advantages not only over classical algorithms but also over previous quantum algorithms. The answer to the question is not obvious. For typical projection-based quantum algorithms, since we don't know how to prepare a good initial state having a large overlap with $|\rho_{ss}\rangle$, it seems $\rho_{ss}$ might be computationally easier to prepare. However, since LME under the vectorization is essentially a linear ordinary differential equation (ODE), we can thus use quantum ODE algorithms \cite{berry2017quantum,childs2020quantum,krovi2023improved,berry2024quantum,fang2023time,an2023linear,jin2022quantum,an2023quantum} to prepare $|\rho_{ss}\rangle$. If we don't pursue using amplitude amplification \cite{brassard2002quantum}, the complexity of these ODE algorithms has $\mathcal{O}(Tr(\rho_{ss}^2)^{-1})$ dependence on the purity. If the task is to give a $\epsilon$-additive estimation of $\langle\rho_{ss}|A|\rho_{ss}\rangle$, both our algorithm and quantum ODE algorithms have a complexity proportional to $\mathcal{O}(Tr(\rho_{ss}^2)^{-1}\epsilon^{-2})$. Thus, our algorithm gives no advantages.

However, we can consider another task of measuring amplitudes of the form $\langle b_i|\rho_{ss}\rangle$ where $|b_i\rangle$ is a $2n$-qubit Bell state (a maximally entangled state between row and column subsystems). Under the vectorization picture, this amplitude corresponds to:
\begin{equation}\label{dfgh}
\langle b_i|\rho_{ss}\rangle=\frac{1}{2^{n/2}\sqrt{Tr(\rho_{ss}^2)}}Tr(P_i\rho_{ss})
\end{equation}
with $P_i$ is a $n$-qubit Pauli operator \cite{shang2024estimating}. On the left-hand side of Eq. \ref{dfgh}, $\langle b_i|\rho_{ss}\rangle$ is a measurable value for $|\rho_{ss}\rangle$ by the Hadamard test (or the amplitude estimation). On the right-hand side, $Tr(P_i\rho_{ss})$ is also a measurable value for $\rho_{ss}$ again by the Hadamard test. To estimate for example $Re[\langle b_i|\rho_{ss}\rangle]$ to an additive error $\epsilon$, we require $\mathcal{O}(\epsilon^{-2})$ queries to the preparation of $|\rho_{ss}\rangle$, resulting in a total complexity of order $\mathcal{O}(Tr(\rho_{ss}^2)^{-1}\epsilon^{-2})$. In contrast, to realize the same accuracy, under the vectorization picture, we only need to give a $2^{n/2}\sqrt{Tr(\rho_{ss}^2)}\epsilon$-additive estimation on $Tr(P_i\rho_{ss})$, resulting in a total complexity of order $\mathcal{O}(2^{-n}Tr(\rho_{ss}^2)^{-1}\epsilon^{-2})$. Thus, for this special task, we can have a $2^n$ complexity reduction using the vectorization picture.

\section{Properties}
Having shown the frame of our algorithm, now we give discussions on several important aspects of our algorithm.

\subsection{Properties of $L^\dag L$}
Firstly, we want to discuss the properties that $L^\dag L$ satisfies under a few very general assumptions, which may give some criteria to tell whether a Hamiltonian can be solved by our algorithm. We summarized two properties of $L^\dag L$ below:
\begin{itemize}[leftmargin=*,itemsep=2pt,topsep=0pt,parsep=0pt]
\item[a] The spectra of $L^\dag L$ is non-negative with the ground state energy equal to zero.
\item[b] $L^\dag L$ has the exchange time reversal symmetry i.e. invariance with respect to the simultaneous action of the exchange operator $S$ between the row subsystem and the column subsystem and time reversal operator $T$ ($T|\psi\rangle=|\psi^*\rangle$): $[L^\dag L, ST]=0$.
\end{itemize}
The above properties come from the Liouvillian operator $L$ of the form Eq.(\ref{3}). Eq.(\ref{3}) ensures that during the LME evolution, $\rho(t)$ is always Hermitian, positive semi-definite and $Tr(\rho(t))=1$. These constraints thus lead to properties of $L$ \cite{minganti2018spectral} and some of them are inherited by $L^\dag L$.

\subsection{Generalizations to other types of Hamiltonian}
Secondly, the Hamiltonians that are solvable by this algorithm not only include $L^\dag L$ but also other types. One natural type is $-L$ when $L$ is Hermitian. In this case, LME induces a pure imaginary time evolution. Another type is to generalize the $L^\dag L$ to its polynomial functions with even parity whose ground states also correspond to the steady state of $L$. 

There is also a type inspired by the idea in Ref. \cite{shang2021schr} where any Hamiltonians that can be transformed into the above types by Clifford circuits can also be efficiently solved by this algorithm i.e. $H=U_cL^\dag LU_c^\dag$. The ground state of $H$ is $U_c|\rho_{ss}\rangle$. While $U_c|\rho_{ss}\rangle$ can't be represented by a density matrix through the mapping Eq. \ref{4} in general, we can just transfer the $U_c$ to $A$ as $A'=U_c^\dag A U_c$ and we have $\langle \rho_{ss}|U_c^\dag AU_c|\rho_{ss}\rangle=\langle \rho_{ss}|A'|\rho_{ss}\rangle$ which can be measured by Eq. \ref{5}. This type may greatly enlarge the range of solvable problems by our algorithm since $U_c|\rho_{ss}\rangle$ can go beyond the restrictions of $|\rho\rangle$ and has been proved to have great expressibility of states \cite{shang2021schr}. 

Moreover, if we don't restrict to the ground state, but eigenstates with certain energy, more types can be constructed. For example, we can define Hamiltonian \cite{ramusat2021quantum} with the form:
\begin{equation}\label{fff}
H=\begin{pmatrix}
0 & L \\
L^\dag & 0
\end{pmatrix}
\end{equation}
Then $|\rho_{ss}\rangle$ is the eigenstate of $H$ with zero energy but not ground energy. The advantage of this construction is that we can avoid the classical XL algorithm part for quadratic equations.

More generally, for any Hamiltonian whose ground state has a polynomial small overlap with $\rho_{ss}$, we can always use our algorithm as an initial state preparation strategy for those projection-based quantum algorithms to efficiently solve the ground state.

\subsection{Emerging "non-linearity``}
Thirdlyly, quantum mechanics is linear regardless of the closed dynamics governed by the Schrödinger equation or the open dynamics governed by the LME. However, interestingly, non-linear dynamics can emerge from such a linear nature after the mapping Eq. \ref{4}. To see this non-linearity, consider three density matrices $\rho_1$, $\rho_2$ and $\rho_3$ where $ \rho_3$ can be expressed as a linear combination of $\rho_1$ and $\rho_2$: $\rho_3=c_1\rho_1+c_2\rho_2$. Using the mapping Eq. \ref{4}, we can have the corresponding DM-states $|\rho_1\rangle$, $|\rho_2\rangle$ and $|\rho_3\rangle$ where $C_{\rho_3}|\rho_3\rangle=c1C_{\rho_1}|\rho_1\rangle+c2C_{\rho_2}|\rho_2\rangle$. Now consider a linear quantum operation $K_D$ which transforms each $\rho_i$ into a $\rho_i'$, then due to the linearity of $K_D$, $\rho_3'=c_1\rho_1'+c_2\rho_2'$. The corresponding DM-states of $\rho_i'$ then have the relation $C_{\rho_3'}|\rho_3'\rangle=c1C_{\rho_1'}|\rho_1'\rangle+c2C_{\rho_2'}|\rho_2'\rangle$. We can define $K_I$ as the corresponding virtual quantum operation of $K_D$ after $f_1$ as $K_I[|\rho\rangle]=f_1[K_D[\rho]]=|\rho'\rangle$, then:
\begin{eqnarray}\label{y4}
&&K_I[\frac{c1C_{\rho_1}}{C_{\rho_3}}|\rho_1\rangle+\frac{c2C_{\rho_2}}{C_{\rho_3}}|\rho_2\rangle]=\nonumber\\&&\frac{c1C_{\rho_1'}}{C_{\rho_3'}}K_I[|\rho_1\rangle]+\frac{c2C_{\rho_2'}}{C_{\rho_3'}}K_I[|\rho_2\rangle]
\end{eqnarray}
Hence, in general cases, $K_I$ is not a linear operator despite the linearity of $K_D$. As shown in Ref. \cite{abrams1998nonlinear}, non-linear quantum mechanics can solve problems that can't be solved efficiently by standard linear quantum mechanics. Thus, this might be an interesting feature to be further investigated.

\section{Summary and outlook}
In summary, we propose a novel dissipation-based quantum algorithm for solving the ground state of a class of Hamiltonians. Here, the word "solving`` means estimating the expectation value of operators with respect to the ground state. When the system is driven to its steady state $\rho_{ss}$, we understand it under the vectorization picture and propose a novel measurement protocol Eq. \ref{5} to obtain information of $|\rho_{ss}\rangle$ from $\rho_{ss}$. We also gave an efficient classical procedure to obtain $L$ from a given Hamiltonian $H$. The quantum easiness regime of our algorithm contains classically hard instances. Compared with previous quantum algorithms without utilizing vectorization, our algorithm can have exponential complexity reductions for Bell basis amplitude estimation tasks.

This algorithm still leaves plenty of exciting questions for future investigations. For instance, it is interesting to think about whether there are more general types of Hamiltonian amenable to our algorithm. Future work on finding Hamiltonian with physical significance and/or practical value to solve using our algorithm is highly demanded. It would be also very attractive to find separations between with and without vectorization for other measurement tasks. What's more, the techniques developed in this work may help the study of non-Hermitian Hamiltonian physics \cite{el2018non,gong2018topological} on quantum computers as Eq. \ref{2} is a natural non-Hermitian evolution and Eq. \ref{5} shows the way for extracting information and preparation. We hope this work will invoke future novel research on quantum algorithms and quantum information science.

See the Appendix for all the details.

\begin{acknowledgments}
This work is supported by the National Natural Science Foundation of China (No. 91836303 and No. 11805197), the National Key R$\&$D Program of China, the Chinese Academy of Sciences, the Anhui Initiative in Quantum Information Technologies, and the Science and Technology Commission of Shanghai Municipality (2019SHZDZX01). The authors would like to thank JI Cirac, X Yuan, and L Lin for their fruitful comments.
\end{acknowledgments}
\bibliographystyle{unsrt}
\bibliography{ref.bib}
\onecolumngrid
\appendix   

\section{Derivations of Eq.(\ref{5})}
$\langle\rho|A|\rho\rangle$ can be measured from $\rho$:
\begin{eqnarray}\label{a1}
\langle\rho|A|\rho\rangle&&=\sum_{ijkl}\frac{1}{C_\rho^2}\rho^*_{ij}A_{ij,kl}\rho_{kl}=\sum_{ijkl}\frac{1}{Tr(\rho^2)}A_{ij,kl}\rho_{ji}\rho_{kl}
\nonumber\\&&=\sum_{ijkl}\frac{1}{Tr(\rho^2)}B_{il,jk}\rho_{ji}\rho_{kl}
=\frac{Tr(B\rho\otimes\rho)}{Tr(\rho^2)}
\end{eqnarray}
where $B$ follow the rules in Eq.(\ref{6}).

\section{2-qubit Pauli substitute operators}

Details of transformations from 2-qubit Pauli operators ($A$) to 2-qubit Pauli substitute operators ($B$) are shown in Table 1-4 where each $B$'s corresponding exact matrix and equivalent quantum circuit for Hadamard test are presented. The only required 2-qubit gate resource is the swap gate.
\begin{table*}[ht]
  \centering
  \begin{tabular}{|c|c|p{0.42\textwidth}<{\centering}|c|}\hline
      ID&$A$ (Hermite)& $B$ (Unitary)& Eigenvalues of B \\\hline
      1&$II$& $0.5 II+0.5XX+0.5YY+0.5ZZ$& diag(1,1,1,-1)\\\hline
      2&$XX$& $0.5 II+0.5XX-0.5YY-0.5ZZ$& diag(1,1,-1,1)\\\hline 
      3&$YY$& $-0.5 II+0.5XX-0.5YY+0.5ZZ$& diag(1,-1,-1,-1)\\\hline
      4&$ZZ$& $0.5 II-0.5XX-0.5YY+0.5ZZ$& diag(1,-1,1,1)\\\hline 
      \multicolumn{4}{|c|}{B matrix}\\\hline
      \multicolumn{4}{|c|}{
      \begin{tabular}{@{} p{0.2\textwidth}|p{0.2\textwidth}|p{0.2\textwidth}|p{0.2\textwidth} @{}}
          1 & 2& 3& 4
      \end{tabular}} 
      \\
      \multicolumn{4}{|c|}{
      \begin{tabular}{@{} p{0.2\textwidth}<{\centering}|p{0.2\textwidth}<{\centering}|p{0.2\textwidth}<{\centering}|p{0.2\textwidth}<{\centering} @{}}
          $\begin{pmatrix}[1]
              1 & 0 & 0 & 0\\
              0 & 0 & 1 & 0\\
              0 & 1 & 0 & 0\\
              0 & 0 & 0 & 1
          \end{pmatrix}$&
          $\begin{pmatrix}[1]
              0 & 0 & 0 & 1\\
              0 & 1 & 0 & 0\\
              0 & 0 & 1 & 0\\
              1 & 0 & 0 & 0
          \end{pmatrix}$&
          $\begin{pmatrix}[1]
              0 & 0 & 0 & 1\\
              0 & -1 & 0 & 0\\
              0 & 0 & -1 & 0\\
              1 & 0 & 0 & 0
          \end{pmatrix}$&
          $\begin{pmatrix}[1]
              1 & 0 & 0 & 0\\
              0 & 0 & -1 & 0\\
              0 & -1 & 0 & 0\\
              0 & 0 & 0 & 1
          \end{pmatrix}$\\
          & & & 
      \end{tabular}} 
      \\\hline
      \multicolumn{4}{|c|}{B circuit}\\\hline
      \multicolumn{4}{|c|}{
      \begin{tabular}{@{} p{0.2\textwidth}|p{0.2\textwidth}|p{0.2\textwidth}|p{0.2\textwidth} @{}}
          1 & 2& 3& 4
      \end{tabular}} 
      \\
      \multicolumn{4}{|c|}{
      \begin{tabular}{@{} p{0.2\textwidth}<{\centering}|p{0.2\textwidth}<{\centering}|p{0.2\textwidth}<{\centering}|p{0.2\textwidth}<{\centering} @{}}
          \
          \Qcircuit @C=3.5em @R=3.5em {
              \lstick{} & \qswap  & \qw \\
              \lstick{} & \qswap \qwx & \qw 
          }&
          \
          \Qcircuit @C=2em @R=2em {
              \lstick{} & \gate{X} & \qswap  & \qw \\
              \lstick{} & \gate{X} & \qswap \qwx & \qw 
          }&
          \
          \Qcircuit @C=2em @R=2em {
              \lstick{} & \gate{i Y} & \qswap  & \qw \\
              \lstick{} & \gate{i Y} & \qswap \qwx & \qw 
          }&
          \
          \Qcircuit @C=2em @R=2em {
              \lstick{} & \gate{Z} & \qswap  & \qw \\
              \lstick{} & \gate{Z} & \qswap \qwx & \qw 
          }\\
          & & &
      \end{tabular}} 
      \\\hline
  \end{tabular}
  \caption{\textbf{Set 1}}
\end{table*}
\begin{table*}[b]
  \centering
  \begin{tabular}{|c|c|p{0.42\textwidth}<{\centering}|c|}\hline
      ID&$A$ (Hermite)& $B$ (Unitary)& Eigenvalues of B \\\hline
      5&$IX$& $0.5 IX+0.5XI+0.5iYZ-0.5iZY$& \{-1,$i$,-$i$,1\}\\\hline 
      6&$XI$& $0.5 IX+0.5XI-0.5iYZ+0.5iZY$& \{-1,$i$,-$i$,1\}\\\hline 
      7&$YZ$& $-0.5i IX+0.5iXI+0.5YZ+0.5ZY$& \{-1,$i$,-$i$,1\}\\\hline
      8&$ZY$& $-0.5i IX+0.5iXI-0.5YZ-0.5ZY$&\{-1,$i$,-$i$,1\}\\\hline
      \multicolumn{4}{|c|}{B matrix}\\\hline
      \multicolumn{4}{|c|}{
      \begin{tabular}{@{} p{0.2\textwidth}|p{0.2\textwidth}|p{0.2\textwidth}|p{0.2\textwidth} @{}}
          5 & 6& 7& 8
      \end{tabular}} 
      \\
      \multicolumn{4}{|c|}{
      \begin{tabular}{@{} p{0.2\textwidth}<{\centering}|p{0.2\textwidth}<{\centering}|p{0.2\textwidth}<{\centering}|p{0.2\textwidth}<{\centering} @{}}
          $\begin{pmatrix}[1]
              0 & 0 & 1 & 0\\
              1 & 0 & 0 & 0\\
              0 & 0 & 0 & 1\\
              0 & 1 & 0 & 0
          \end{pmatrix}$&
          $\begin{pmatrix}[1]
              0 & 1 & 0 & 0\\
              0 & 0 & 0 & 1\\
              1 & 0 & 0 & 0\\
              0 & 0 & 1 & 0
          \end{pmatrix}$&
          $\begin{pmatrix}[1]
              0 & -i & 0 & 0\\
              0 & 0 & 0 & i\\
              i & 0 & 0 & 0\\
              0 & 0 & -i & 0
          \end{pmatrix}$&
          $\begin{pmatrix}[1]
              0 & 0 & i & 0\\
              -i & 0 & 0 & 0\\
              0 & 0 & 0 & -i\\
              0 & i & 0 & 0
          \end{pmatrix}$\\
          & & & 
      \end{tabular}} 
      \\\hline
      \multicolumn{4}{|c|}{B circuit}\\\hline
      \multicolumn{4}{|c|}{
      \begin{tabular}{@{} p{0.2\textwidth}|p{0.2\textwidth}|p{0.2\textwidth}|p{0.2\textwidth} @{}}
          5 & 6& 7& 8
      \end{tabular}} 
      \\
      \multicolumn{4}{|c|}{
      \begin{tabular}{@{} p{0.2\textwidth}<{\centering}|p{0.2\textwidth}<{\centering}|p{0.2\textwidth}<{\centering}|p{0.2\textwidth}<{\centering} @{}}
          \
          \Qcircuit @C=2.2em @R=2.2em {
              \lstick{} & \gate{X} & \qswap  & \qw \\
              \lstick{} & \qw & \qswap \qwx & \qw 
          }&
          \
          \Qcircuit @C=2.2em @R=2.2em {
              \lstick{} & \qw & \qswap  & \qw \\
              \lstick{} & \gate{X} & \qswap \qwx & \qw 
          }&
          \
          \Qcircuit @C=2em @R=2em {
              \lstick{} & \gate{Z} & \qswap  & \qw \\
              \lstick{} & \gate{Y} & \qswap \qwx & \qw 
          }&
          \
          \Qcircuit @C=2em @R=2em {
              \lstick{} & \gate{i Y} & \qswap  & \qw \\
              \lstick{} & \gate{i Z} & \qswap \qwx & \qw 
          }\\
          & & & 
      \end{tabular}} 
      \\\hline
  \end{tabular}
  \caption{\textbf{Set 2}}
\end{table*}
\clearpage
\begin{table*}[t]   
  \centering 
  \begin{tabular}{|c|c|p{0.42\textwidth}<{\centering}|c|}\hline
      ID&$A$ (Hermite)& $B$ (Unitary)& Eigenvalues of B \\\hline
      9&$IY$& $-0.5 IY+0.5iXZ-0.5YI-0.5iZX$& \{-1,$i$,-$i$,1\}\\\hline 
      10&$YI$& $0.5 IY+0.5iXZ+0.5YI-0.5iZX$& \{-1,$i$,-$i$,1\}\\\hline 
      11&$XZ$& $0.5i IY+0.5XZ-0.5iYI+0.5ZX$& \{-1,$i$,-$i$,1\}\\\hline 
      12&$ZX$& $-0.5i IY+0.5XZ+0.5iYI+0.5ZX$& \{-1,$i$,-$i$,1\}\\\hline  
      \multicolumn{4}{|c|}{B matrix}\\\hline
      \multicolumn{4}{|c|}{
      \begin{tabular}{@{} p{0.2\textwidth}|p{0.2\textwidth}|p{0.2\textwidth}|p{0.2\textwidth} @{}}
          9 & 10& 11& 12
      \end{tabular}} 
      \\
      \multicolumn{4}{|c|}{
      \begin{tabular}{@{} p{0.2\textwidth}<{\centering}|p{0.2\textwidth}<{\centering}|p{0.2\textwidth}<{\centering}|p{0.2\textwidth}<{\centering} @{}}
          $\begin{pmatrix}[1]
              0 & 0 & i & 0\\
              -i & 0 & 0 & 0\\
              0 & 0 & 0 & i\\
              0 & -i & 0 & 0
          \end{pmatrix}$&
          $\begin{pmatrix}[1]
              0 & -i & 0 & 0\\
              0 & 0 & 0 & -i\\
              i & 0 & 0 & 0\\
              0 & 0 & i & 0
          \end{pmatrix}$&
          $\begin{pmatrix}[1]
              0 & 1 & 0 & 0\\
              0 & 0 & 0 & -1\\
              1 & 0 & 0 & 0\\
              0 & 0 & -1 & 0
          \end{pmatrix}$&
          $\begin{pmatrix}[1]
              0 & 0 & 1 & 0\\
              1 & 0 & 0 & 0\\
              0 & 0 & 0 & -1\\
              0 & -1 & 0 & 0
          \end{pmatrix}$\\
          & & & 
      \end{tabular}} 
      \\\hline
      \multicolumn{4}{|c|}{B circuit}\\\hline
      \multicolumn{4}{|c|}{
      \begin{tabular}{@{} p{0.2\textwidth}|p{0.2\textwidth}|p{0.2\textwidth}|p{0.2\textwidth} @{}}
          9 & 10& 11& 12
      \end{tabular}} 
      \\
      \multicolumn{4}{|c|}{
      \begin{tabular}{@{} p{0.2\textwidth}<{\centering}|p{0.2\textwidth}<{\centering}|p{0.2\textwidth}<{\centering}|p{0.2\textwidth}<{\centering} @{}}
          \
          \Qcircuit @C=2.2em @R=2.2em {
              \lstick{} & \gate{-Y} & \qswap  & \qw \\
              \lstick{} & \qw & \qswap \qwx & \qw 
          }&
          \
          \Qcircuit @C=2.2em @R=2.2em {
              \lstick{} & \qw & \qswap  & \qw \\
              \lstick{} & \gate{Y} & \qswap \qwx & \qw 
          }&
          \
          \Qcircuit @C=2em @R=2em {
              \lstick{} & \gate{Z} & \qswap  & \qw \\
              \lstick{} & \gate{X} & \qswap \qwx & \qw 
          }&
          \
          \Qcircuit @C=2em @R=2em {
              \lstick{} & \gate{X} & \qswap  & \qw \\
              \lstick{} & \gate{Z} & \qswap \qwx & \qw 
          }\\
          & & & 
      \end{tabular}} 
      \\\hline
  \end{tabular}
  \caption{\textbf{Set 3}}
\end{table*}
\begin{table*}[b]
  \centering
  \begin{tabular}{|c|c|p{0.42\textwidth}<{\centering}|c|}\hline
      ID&$A$ (Hermite)& $B$ (Unitary)& Eigenvalues of B \\\hline
      13&$IZ$& $0.5 IZ+0.5iXY-0.5iYX+0.5ZI$& \{$i$,-$i$,1,-1\}\\\hline 
      14&$ZI$& $0.5 IZ-0.5iXY+0.5iYX+0.5ZI$& \{$i$,-$i$,1,-1\}\\\hline   
      15&$XY$& $0.5i IZ-0.5XY-0.5YX-0.5iZI$& \{1,-1,-$i$,$i$\}\\\hline
      16&$YX$& $0.5i IZ+0.5XY+0.5YX-0.5iZI$& \{1,-1,-$i$,$i$\}\\\hline
      \multicolumn{4}{|c|}{B matrix}\\\hline
      \multicolumn{4}{|c|}{
      \begin{tabular}{@{} p{0.2\textwidth}|p{0.2\textwidth}|p{0.2\textwidth}|p{0.2\textwidth} @{}}
          13 & 14& 15& 16
      \end{tabular}} 
      \\
      \multicolumn{4}{|c|}{
      \begin{tabular}{@{} p{0.2\textwidth}<{\centering}|p{0.2\textwidth}<{\centering}|p{0.2\textwidth}<{\centering}|p{0.2\textwidth}<{\centering} @{}}
          $\begin{pmatrix}[1]
              1 & 0 & 0 & 0\\
              0 & 0 & -1 & 0\\
              0 & 1 & 0 & 0\\
              0 & 0 & 0 & -1
          \end{pmatrix}$&
          $\begin{pmatrix}[1]
              1 & 0 & 0 & 0\\
              0 & 0 & 1 & 0\\
              0 & -1 & 0 & 0\\
              0 & 0 & 0 & -1
          \end{pmatrix}$&
          $\begin{pmatrix}[1]
              0 & 0 & 0 & i\\
              0 & -i & 0 & 0\\
              0 & 0 & i & 0\\
              -i & 0 & 0 & 0
          \end{pmatrix}$&
          $\begin{pmatrix}[1]
              0 & 0 & 0 & -i\\
              0 & -i & 0 & 0\\
              0 & 0 & i & 0\\
              i & 0 & 0 & 0
          \end{pmatrix}$\\
          & & & 
      \end{tabular}} 
      \\\hline
      \multicolumn{4}{|c|}{B circuit}\\\hline
      \multicolumn{4}{|c|}{
      \begin{tabular}{@{} p{0.2\textwidth}|p{0.2\textwidth}|p{0.2\textwidth}|p{0.2\textwidth} @{}}
          13 & 14& 15& 16
      \end{tabular}} 
      \\
      \multicolumn{4}{|c|}{
      \begin{tabular}{@{} p{0.2\textwidth}<{\centering}|p{0.2\textwidth}<{\centering}|p{0.2\textwidth}<{\centering}|p{0.2\textwidth}<{\centering} @{}}
          \
          \Qcircuit @C=2.2em @R=2.2em {
              \lstick{} & \gate{Z} & \qswap  & \qw \\
              \lstick{} & \qw & \qswap \qwx & \qw 
          }&
          \
          \Qcircuit @C=2.2em @R=2.2em {
              \lstick{} & \qw & \qswap  & \qw \\
              \lstick{} & \gate{Z} & \qswap \qwx & \qw 
          }&
          \
          \Qcircuit @C=2em @R=2em {
              \lstick{} & \gate{iY} & \qswap  & \qw \\
              \lstick{} & \gate{iX} & \qswap \qwx & \qw 
          }&
          \
          \Qcircuit @C=2em @R=2em {
              \lstick{} & \gate{X} & \qswap  & \qw \\
              \lstick{} & \gate{Y} & \qswap \qwx & \qw 
          }\\
          & & & 
      \end{tabular}} 
      \\\hline
  \end{tabular}
  \caption{\textbf{Set 4}}
\end{table*}
\clearpage
\section{The concrete form of $L^\dag L$}
The dynamics of open systems in Lindblad form is described by
\begin{equation}
    \frac{d\rho}{dt} =L[\rho] = -i[H,\rho] + \sum_{i}\lambda_i(F_i\rho F_i^{\dagger}-\frac{1}{2}\{\rho,F_i^{\dagger}F_i\})
\end{equation}
where $L$ can be written in the matrix form as
\begin{equation}
    L = -i(H\otimes I-I\otimes H^*) + \sum_{i}\lambda_i (F_i\otimes F_i^*-\frac{1}{2}F^{\dagger}_iF_i\otimes I-\frac{1}{2}I\otimes F_i^TF_i^*)
\end{equation}
Hence, after some algebra, we can see that $L^{\dagger}L$ takes the following form
\begin{eqnarray}
    L^{\dagger}L =&& H^2\otimes I +I\otimes (H^*)^2 -H\otimes H^*-H\otimes H^*\\
    &&+\sum_{\alpha}\lambda_{\alpha}\left\{ iHF_\alpha\otimes F_\alpha^* -iF_\alpha^{\dagger}H\otimes F_\alpha^T + iF_{\alpha}^{\dagger}\otimes F_{\alpha}^TH^T - iF_{\alpha}\otimes H^*F_{\alpha}^* \right.\\
    &&\qquad \qquad   +\frac{i}{2}[F_{\alpha}^{\dagger}F_\alpha,H]\otimes I + \frac{i}{2}I\otimes[H^*,F^{T}_{\alpha}F_{\alpha}^*] \}\\
    &&+\sum_{\alpha\beta}\lambda_\alpha\lambda_\beta \left\{  
    F_{\alpha}^{\dagger}F_{\beta}\otimes F_{\alpha}^{T}F_{\beta}^* -\frac{1}{2}F_{\alpha}^{\dagger}\otimes F_{\alpha}^TF_{\beta}^TF_{\beta}^*  -\frac{1}{2}F_{\beta}\otimes F_{\alpha}^TF_{\alpha}^*F_{\beta}^*
    \right.\\
    &&\qquad \qquad \quad -\frac{1}{2}F_{\alpha}^{\dagger}F_{\beta}^{\dagger}F_{\beta}\otimes F_{\alpha}^T-\frac{1}{2}F_{\alpha}^{\dagger}F_{\alpha}F_{\beta}\otimes F_{\beta}^*\\
    &&\qquad \qquad \quad +\frac{1}{4}F_{\alpha}^{\dagger}F_{\alpha}\otimes F_{\beta}^{T}F_{\beta}^* +\frac{1}{4}F_{\beta}^{\dagger}F_{\beta}\otimes F_{\alpha}^TF_{\alpha}^*\\
    &&\qquad \qquad \quad
    \left. +\frac{1}{4}I\otimes F_{\alpha}^TF_{\alpha}^*F_{\beta}^TF_{\beta}^* +\frac{1}{4}F_{\alpha}^{\dagger}F_{\alpha}F_{\beta}^{\dagger}F_{\beta}\otimes I      \right\}
\end{eqnarray}

\section{Measurement runtime of Eq.(\ref{5})}\label{sec: Measurement complexity 1}
Before the analysis, we will introduce the estimator that will be used. 
\begin{itemize}
	\item Estimator for $\frac{E[X]}{E[Y]}$: When $X$ and $Y$ are two independent random variables, the value of the ratio of their expectations $\frac{E[X]}{E[Y]}$ can be estimated by an asymptotically unbiased estimator $\frac{\overline{X}}{\overline{Y}}$ where $\overline{X}$ and $\overline{Y}$ are the averages of $X$ and $Y$. It has been shown in Ref.\cite{van2000mean} that the expectation and the variance of this estimator are:
    \begin{eqnarray}
    E[\frac{\overline{X}}{\overline{Y}}]&&\approx \frac{E[X]}{E[Y]}+\frac{E[X]}{E[Y]^3}Var[\overline{Y}]\label{m1}\\
    Var[\frac{\overline{X}}{\overline{Y}}]&&\approx \frac{E[X]^2Var[\overline{Y}]+E[Y]^2Var[\overline{X}]}{E[Y]^4}\label{m2}
    \end{eqnarray}
\end{itemize}
Eq.(\ref{m1}) and Eq.(\ref{m2}) are good approximations when the uncertainties of the averages are small enough which in our case corresponds to enough measurement times.

In the following, without loss of generality, we will assume $A$ is Hermite and has the same form in the main text:
\begin{equation}\label{m7}
    A=\sum_{i=1}^m g_iP_i
\end{equation}
One can easily generalize the results of this section to non-Hermite $A$ cases.

We now analyze the measurement complexity of $\langle \rho|A|\rho\rangle=\frac{Tr(B\rho\otimes\rho)}{Tr(\rho^2)}$. Each $Re(Tr(Q_i\rho\otimes\rho))$ in $Tr(B\rho\otimes\rho)$ is the expectation of a Bernoulli random variable $Z_i$: $E[Z_i]=p_i*1+(1-p_i)*(-1)$ with a variance $Var[Z_i]=4p_i (1-p_i)\leq 1$. Suppose each $Q_i$ is measured for $\frac{N_h}{m}$ times to obtain the average of the numerator $\overline{Tr(B\rho\otimes\rho)}$, then the total variance of evaluating $Tr(B\rho\otimes\rho)=\sum_{i=1}^m g_i(2p_i-1)$ is:
\begin{equation}\label{m8}
    Var[\overline{Tr(B\rho\otimes\rho)}]=\sum_{i=1}^m \frac{mg_i^24p_i (1-p_i)}{N_h}
\end{equation}
The purity $Tr(\rho^2)$ is also the expectation of a Bernoulli random variable $Z_s$ and we assume it has a lower bound: $E[Z_s]=p_s*1+(1-p_s)*(-1)\geq \gamma$ with a variance $Var[Z_i]=4p_i (1-p_i)\leq 1$. Then the variance of average $\overline{Tr(\rho^2)}$ for $N_s$ measurement times on the purity $Tr(\rho^2)=2p_s-1$ is:
\begin{equation}\label{m9}
    Var[\overline{Tr(\rho^2)}]=\frac{4p_s(1-p_s)}{N_s}
\end{equation}
According to Eq.(\ref{m1}) and Eq.(\ref{m2}), when $\frac{\overline{Tr(B\rho\otimes\rho)}}{\overline{Tr(\rho^2)}}$ are used to estimate $\langle \rho|A|\rho\rangle$, the bias is:
\begin{eqnarray}\label{m10}
Bias[\langle \rho|A|\rho\rangle]^2&&=\left(\frac{\overline{Tr(B\rho\otimes\rho)}}{\overline{Tr(\rho^2)}}-\langle \rho|A|\rho\rangle\right)^2\nonumber\approx\left(\frac{Tr(B\rho\otimes\rho)}{Tr(\rho^2)^3}Var[\overline{Tr(\rho^2)}]\right)^2
\nonumber\\&& =\left(\frac{\sum_{i=1}^m g_i(2p_i-1)}{(2p_s-1)^3}\frac{4p_s(1-p_s)}{N_s}\right)^2=\left(\frac{\langle \rho|A|\rho\rangle}{(2p_s-1)^2}\frac{4p_s(1-p_s)}{N_s}\right)^2\nonumber\\&& \leq \frac{\langle \rho|A|\rho\rangle^2}{(2p_s-1)^4N_s^2} \leq \frac{\langle \rho|A|\rho\rangle^2}{\gamma^4 N_s^2}\leq \frac{||A||_2^2}{\gamma^4 N_s^2}
\end{eqnarray}
and the variance is:
\begin{eqnarray}\label{m11}
Var[\langle \rho|A|\rho\rangle]&&\approx\frac{E[Tr(B\rho\otimes\rho)]^2Var[\overline{Tr(\rho^2)}]+Tr(\rho^2)^2Var[\overline{Tr(B\rho\otimes\rho)}]}{Tr(\rho^2)^4}\nonumber\\&&=\frac{(\sum_{i=1}^m g_i(2p_i-1))^2\frac{4p_s(1-p_s)}{N_s}+(2p_s-1)^2\sum_{i=1}^m \frac{mg_i^24p_i (1-p_i)}{N_h}}{(2p_s-1)^4}\nonumber\\&&=\frac{\langle \rho|A|\rho\rangle^2\frac{4p_s(1-p_s)}{N_s}+\sum_{i=1}^m \frac{mg_i^24p_i (1-p_i)}{N_h}}{(2p_s-1)^2}\nonumber\\&&\leq \frac{\langle \rho|A|\rho\rangle^2}{\gamma^2N_s}+\frac{\sum_{i=1}^m mg_i^2}{\gamma^2N_h}\leq\frac{||A||_2^2}{\gamma^2N_s}+\frac{m||A||_F^2}{2^{2n}\gamma^2N_h}
\end{eqnarray}
where $||A||_2$ and $||A||_F$ are the 2-norm and the Frobenius-norm of $A$. 
Thus we obtain the mean squared error (MSE) \cite{james2013introduction} for measuring $\langle \rho|A|\rho\rangle$:
\begin{eqnarray}\label{m12}
MSE[\langle \rho|A|\rho\rangle]&&=Bias[\langle \rho|A|\rho\rangle]^2+Var[\langle \rho|A|\rho\rangle]\nonumber\\&&\leq\frac{||A||_2^2}{\gamma^4 N_s^2}+\frac{||A||_2^2}{\gamma^2N_s}+\frac{m||A||_F^2}{2^{2n}\gamma^2N_h}
\approx \frac{||A||_2^2}{\gamma^2N_s}+\frac{m||A||_F^2}{2^{2n}\gamma^2N_h}
\end{eqnarray}
We can set $N_s=N_h=\frac{N}{2}$, then the total number of measurements $N$ needed to achieve an accuracy $\epsilon$ is:
\begin{eqnarray}\label{m13}
N\geq\frac{2}{\gamma^2\epsilon^2}\left(||A||_2^2+\frac{m||A||_F^2}{2^{2n}} \right)
\end{eqnarray}
From Eq.(\ref{m13}), we can conclude that the measurement of $\langle \rho|A|\rho\rangle$ is efficient for $Tr(\rho^2)$ with an lower bound. 

\section{Runtime of LME evolution}
In this section, we will analysis the runtime of LME evolution.

Consider an open quantum system with an initial state $\rho_0$ whose evolution is governed by an LME with corresponding Liouvillian operator $L$. In the following, $\vec{\rho}$ denotes the vector form of $\rho$ and $|\rho\rangle$ denotes the normalized $\vec{\rho}$. We use $\rho_t$ to represent the state of the system at time $t$. For arbitrary $\vec{\rho_0}$, we can decompose it in terms of right eigenvectors of $L$ assuming $L$ is diagonalizable \cite{minganti2018spectral}:
\begin{eqnarray}\label{l1}
\vec{\rho_0}=c_0|r_0\rangle+\sum_{\alpha=1} c_\alpha|r_\alpha\rangle
\end{eqnarray}
where $|r_\alpha\rangle$ are normalized left eigenvectors satisfying $L|r_\alpha\rangle=\eta_\alpha |r_\alpha\rangle$ with $Re[\eta_\alpha]\leq 0$. $|r_0\rangle$ with $\eta_0=0$ is exactly the $|\rho_{ss}\rangle$ and $c_0|r_0\rangle=\vec{\rho_{ss}}$. For the simplicity of the below analysis, we will set $c_\alpha$ to be real and put the phase freedoms into the basis $|r_\alpha\rangle$. For $\vec{\rho_t}$, we can similarly write down its decomposition by solving the LME $d\vec{\rho}/dt=L\vec{\rho}$:
\begin{eqnarray}\label{l2}
\vec{\rho_t}=c_0|r_0\rangle+\sum_{\alpha=1} c_\alpha e^{\eta_\alpha t}|r_\alpha\rangle
\end{eqnarray}
From Eq.(\ref{l2}), we can see that since $Re[\eta_\alpha]<0$ for $\alpha>0$, the populations of $\vec{\rho_t}$ on $|r_\alpha\rangle$ are exponentially decaying except for the steady state.

We will first consider the cases where $L$ is Hermitian. Since $L$ is Hermitian, we have $\langle r_\alpha|r_\beta\rangle=\delta_{\alpha \beta}$. For $|\rho_0\rangle$, its overlap $\zeta$ with $|\rho_{ss}\rangle$ can be defined and calculated:
\begin{eqnarray}\label{l3}
\zeta=|\langle \rho_{ss}|\rho_0\rangle|=\sqrt{\frac{c_0^2}{c_0^2+\sum_{\alpha=1}c_\alpha^2}}
\end{eqnarray}
We can define $X=c_0^2$ and $Y=\sum_{\alpha=1}c_\alpha^2$, then we have:
\begin{eqnarray}\label{l4}
\zeta=\sqrt{\frac{X}{X+Y}}
\end{eqnarray}
For $|\rho_t\rangle$, its overlap with $|\rho_{ss}\rangle$ can be similarly calculated:
\begin{eqnarray}\label{l5}
|\langle \rho_{ss}|\rho_t\rangle|=\sqrt{\frac{c_0^2}{c_0^2+\sum_{\alpha=1}c_\alpha^2 e^{2Re[\eta_\alpha] t}}} >\sqrt{\frac{X}{X+Ye^{-2\Delta_l t}}}
\end{eqnarray}
where we define the asymptotic decay rate $\Delta_l=min(|Re[\eta_\alpha]|)$ which is the smallest real part of the spectra gaps between the steady state and other eigenstates. If we want to reach an overlap $|\langle \rho_{ss}|\rho_t\rangle| >1-\varepsilon $, we can require:
\begin{eqnarray}\label{l6}
\sqrt{\frac{X}{X+Ye^{-2\Delta_l t}}}>1-\varepsilon
\end{eqnarray}
By combining Eq.(\ref{l4}) and the fact $0<\varepsilon<1$ with Eq.(\ref{l6}), we can further require the following inequality holds:
\begin{eqnarray}\label{l7}
\sqrt{\frac{1}{1+\frac{1-\zeta^2}{\zeta^2}e^{-2\Delta_l t}}}>1-\varepsilon
\end{eqnarray}
which leads to:
\begin{eqnarray}\label{l8}
e^{2\Delta_l t}>\varepsilon^{-1}\zeta^{-2}>\frac{1-\varepsilon}{\varepsilon}\frac{1-\zeta^{2}}{\zeta^{2}}
\end{eqnarray}
Thus the runtime dependence on $\Delta_l$, $\zeta$ and $\varepsilon$ follows:
\begin{eqnarray}\label{l9}
t>\Delta_l^{-1}(\ln(\varepsilon^{-\frac{1}{2}})+\ln(\zeta^{-1}))
\end{eqnarray}
The thing is not over here. Eq.(\ref{l3}) can be re-expressed:
\begin{eqnarray}\label{l10}
\zeta=\sqrt{\frac{Tr(\rho_{ss}^2)}{Tr(\rho_0^2)}}
\end{eqnarray}
where we use the relation $\vec{\rho^*}\cdot\vec{\rho}=Tr(\rho^2)$. For the ratio $Tr(\rho_{ss}^2)/Tr(\rho_0^2)$ in Eq.(\ref{l10}), we know the smallest value corresponds to a pure state in the denominator and a maximally mixed state in the numerator. For an $n$-qubit system, we have $Tr(\rho_{ss}^2)/Tr(\rho_0^2)\geq 2^{-n}$. Thus, we have:
\begin{eqnarray}\label{l11}
t>\Delta_l^{-1}(\frac{\ln (2)}{2}n+\ln(\varepsilon^{-\frac{1}{2}}))
\end{eqnarray}

Now we consider the general case where $L$ is not necessarily Hermitian. In this case, $L$ may not even be diagonalizable, thus, we typically use the mixing time $t_{mix}$ to replace the spectral gap $\Delta_l$. The mixing time $t_mix$ is defined as the smallest time such that:
\begin{eqnarray}\label{mix1}
\|e^{Lt_{mix}} (\rho_1-\rho_2)\|_1\leq \frac{1}{2}\|\rho_1-\rho_2\|_1\text{, for any $\rho_1$ and $\rho_2$}
\end{eqnarray}
Based on results in Ref. \cite{minganti2018spectral}, we can set:
\begin{eqnarray}
\vec{\rho_{ss}}&&=\sqrt{Tr(\rho_{ss}^2)}|r_o\rangle\text{, }|\rho_{ss}\rangle=|r_o\rangle\nonumber\\
\vec{\rho_{t}}&&=\sqrt{Tr(\rho_{t}^2)}(c_o|r_o\rangle+c_1|r_1\rangle)\text{, }|\rho_{t}\rangle=c_o|r_o\rangle+c_1|r_1\rangle
\end{eqnarray}
where $c_0$ and $c_1$ are defined as positive numbers such that $c_0^2+c_1^2+c_0c_1(\langle r_0|r_1\rangle+\langle r_1|r_0\rangle)=1$. Then we have:
\begin{eqnarray}
&&|\langle \rho_{ss}|\rho_t\rangle|^2=c_0^2+c_0c_1(\langle r_0|r_1\rangle+\langle r_1|r_0\rangle)+c_1^2 |\langle r_1|r_0\rangle|^2\geq (1-\varepsilon)^2\nonumber\\&&\rightarrow c_1^2(1-|\langle r_1|r_0\rangle|^2)\leq \varepsilon(2-\varepsilon)
\end{eqnarray}
To satisfy this condition, we can simply require:
\begin{eqnarray}\label{xxxx}
c_1^2\leq \varepsilon
\end{eqnarray}
Since we have:
\begin{eqnarray}
\|\rho_t-\rho_{ss}\|_1=\|e^{Lt} (\rho_0-\rho_{ss})\|_1\leq 2^{-\frac{t}{t_{mix}}}\|\rho_0-\rho_{ss}\|_1\leq 2^{-\frac{t}{t_{mix}}}
\end{eqnarray}
where the last inequality is due to $\|\rho_1-\rho_2\|_1\leq 1$ for any $\rho_1$ and $\rho_2$, and we have:
\begin{eqnarray}
\|\rho_t-\rho_{ss}\|_1\geq\|\rho_t-\rho_{ss}\|_2=c_1\sqrt{Tr(\rho_t^2)}
\end{eqnarray}
thus, we obtain the relation:
\begin{eqnarray}\label{xxxxx}
c_1^2\leq \frac{2^{-\frac{2t}{t_{mix}}}}{Tr(\rho_t^2)}\leq 2^{n-\frac{2t}{t_{mix}}}
\end{eqnarray}
Combining Eq. \ref{xxxx} and Eq. \ref{xxxxx}, we get the following scaling:
\begin{eqnarray}\label{xxt}
t\geq \frac{t_{mix}(n+log(\varepsilon^{-1}))}{2}
\end{eqnarray}

In summary, Eq. \ref{l11} and Eq. \ref{xxt} gives the overall runtime bound of LME evolution. Both of which have $\mathcal{O}(n log(\varepsilon^{-1}))$ scaling on the number of qubits and the inaccuracy.

We want to mention that a polynomial small spectral gap of $L$ doesn't necessarily correspond to a polynomial large mixing time \cite{temme2010chi}. For example, when $L$ is non-diagonalizable \cite{minganti2019quantum, rivas2012open, minganti2018spectral}. For nondiagonalizable $L$, the dynamics of $\rho$ can be similarly analyzed using the Jordan canonical form of $L$. A $k\times k$ Jordan block in $L$ corresponding to eigenvalue $\lambda$  which can incur decaying behaviors $f(t)\exp(-|\mathrm{Re}[\lambda]|t)$ where $f(t)$ is a degree-($k-1$) polynomial of $t$. So, for a large system with qubit number $n$ and the corresponding Hilbert space with size $\exp(O(n))$, a Jordan block of dimension $\exp(O(n))$ corresponding to an eigenvalue $\lambda$ with $|\mathrm{Re}[\lambda]|>0$ would result in a relaxation time$\sim \exp(O(n))$ in the worst case.

\section{Properties of $L^\dag L$}
The form of the Liouvillian operator $L$ Eq.(\ref{3}) can preserve the properties of density matrix i.e. the Hermiticity, the positive semi-definiteness and $Tr(\rho)=1$ which will leads to the following properties. 

\romannumeral1: The spectra of $L^\dag L$ is non-negative with the ground energy equals to zero. This property is kind of obvious since the spectra of $L^\dag L$ must be positive except for the ground states with zero energy which correspond to the steady density matrices $\mathcal{L}[\rho_{ss}]=0$. We want to mention here that in order to keep the positive semi-definiteness of density matrix, coefficients $\gamma_i$ need to be positive. This restriction leads to that the real part of the eigenvalues of $L$ are negative. The information of this restriction however, is lost in $L^\dag L$. The reason is that for arbitrary unitary $U$, we have $L^\dag U^\dag U L=L^\dag L$. Thus, the positivity of $\gamma_i$ can be destroyed by $U$.

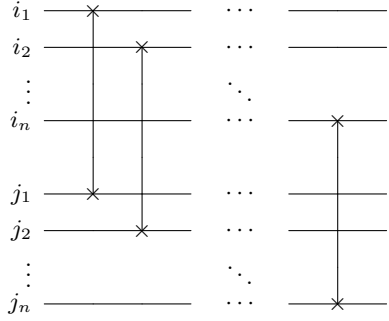
\begin{figure}[htb]
	\centering
    \[ \Qcircuit @C=2em @R=1.5em {
      \lstick{i_1} & \qswap  & \qw & \qw &\cdots  & &\qw & \qw\\
      \lstick{i_2} & \qw \qwx & \qswap & \qw &\cdots & &\qw & \qw\\
      \lstick{\vdots} & \qwx&\qwx &  &\ddots  & & &\\
      \lstick{i_n} & \qw\qwx & \qw\qwx & \qw &\cdots  & &\qswap & \qw\\
      \lstick{} & \qwx & \qwx & & & & \qwx & \\
      \lstick{j_1} & \qswap \qwx & \qw\qwx & \qw &\cdots &  &\qw\qwx & \qw\\
      \lstick{j_2} & \qw & \qswap \qwx & \qw &\cdots &  &\qw\qwx & \qw\\
      \lstick{\vdots} & & &  &\ddots &  &\qwx &\\
      \lstick{j_n}& \qw & \qw & \qw &\cdots &  & \qswap \qwx & \qw
    }\]   
	\caption{The quantum circuit for the exchange operator $S$.}
	\label{fs1}
\end{figure}

\romannumeral2: $L^\dag L$ has the exchange time reversal symmetry i.e. invariance with respect to the simultaneous action of the exchange operator $S$ between the row subsystem and the column subsystem and time reversal operator $T$ ($T|\psi\rangle=|\psi^*\rangle$): $[L^\dag L, ST]=0$. The exchange operator $S$ can be expressed in terms of quantum circuit shown in Fig.(\ref{fs1}) where $i$ indexes denote row subsystem and $j$ indexes denote column subsystem. The exchange time reversal symmetry of $L^\dag L$ comes from the preservation of the Hermiticity of density matrix by $L$ and the action of $ST$ corresponds to the conjugate transpose operation of matrix. In Ref. \cite{minganti2018spectral}, authors proved that assuming the Liouvillian is diagonalizable, if $L|r_\alpha\rangle=e_\alpha|r_\alpha\rangle$, then $L|r_\alpha^\dag\rangle=e_\alpha^*|r_\alpha^\dag\rangle$ where $|r_\alpha\rangle$ are right eigenvectors of $L$. For an arbitrary quantum state $|\psi\rangle$, we can always decompose it into $|\psi\rangle=\sum_\alpha c_\alpha|r_\alpha\rangle$ (which is right except for some exceptional points). Then we have:
\begin{equation}\label{p1}
	(ST)L|\psi\rangle=(ST)\sum_\alpha e_\alpha c_\alpha|r_\alpha\rangle=\sum_\alpha e^*_\alpha c^*_\alpha|r_\alpha^\dag\rangle=L\sum_\alpha c_\alpha^* |r_\alpha^\dag\rangle=L (ST) |\psi\rangle
\end{equation}
Thus, we have $[ST,L]=0$ which is inherited by $L^\dag L$:
\begin{eqnarray}\label{p2}
	\langle\psi|(ST)^\dag L^\dag L (ST)|\psi\rangle &&=\sum_{\alpha\beta}c_\alpha e_\alpha (c_\beta e_\beta)^* \langle r_\alpha^\dag|r_\beta^\dag\rangle \nonumber\\&& =(\sum_{\alpha\beta}(c_\alpha e_\alpha)^* c_\beta e_\beta \langle r_\alpha|S^\dag S|r_\beta\rangle)^*=\langle\psi| L^\dag L |\psi\rangle^* \nonumber\\&& =\langle\psi| L^\dag L |\psi\rangle
\end{eqnarray}

Thus, we can conclude $[L^\dag L, ST]=0$.

For the $Tr(\rho)=1$ property, given a Liouvillian operator $L$, we can expressed it in terms of its eigen basis assuming the Liouvillian is diagonalizable:
\begin{eqnarray}\label{p3}
L=\sum_\alpha e_\alpha |r_\alpha\rangle\langle l_\alpha|
\end{eqnarray}
where $\alpha=0$ corresponds to steady state (i.e. $|r_0\rangle=|\rho_{ss}\rangle$) with $e_0=0$. $\langle l_\alpha|$ and $|r_\alpha\rangle$ are unnormalized left and right eigenvectors of $L$ respectively which satisfy the bi-orthogonal relation:
\begin{eqnarray}\label{p4}
\langle l_\alpha|r_\beta\rangle=\delta_{\alpha\beta}
\end{eqnarray}
In Ref.\cite{minganti2018spectral}, authors proved that the value $\sum_i \langle ii|r_\alpha\rangle$ which corresponds to the trace of matrix equals to 0 if $Re[e_\alpha]\neq 0$. This property is not hold on left eigenvectors because the left eigenvectors of $L$ correspond to right eigenvectors of $L^\dag$ and $L^\dag$ is not a Liouvillian operator. Thus, the $Tr(\rho)=1$ property has no direct indications on $L^\dag L$.

\section{From $L^\dag L$ to $L$ and LME}
The purpose of this appendix is for the situations where we only have Hamiltonians at hands but don't know the corresponding LME. Without any assumptions or restrictions, obtaining $L$ is rather hard since this requires to solve an exponential-multivariate quadratic equations. The complexity of solving general quadratic equations is $\mathcal{O}(exp(N_u))$ with $N_u$ the number of unknowns and has been proved to be NP-hard. However, in our case, we can set some practical constraints to make this problem much easier and be solved in polynomial-time. 

As a specific case, we can constrain the Liouvillian operator $L$ and the Hamiltonian $L^\dag L$ to be local operators. We will call an operator to be $k$-local if every Pauli term of it has {$X$, $Y$, $Z$} act on at most $k$ qubits. Note that this constraint makes very good sense since, on the one hand, local Hamiltonians are of practical interest, on the other hand, it is difficult for quantum computers to run highly non-local LME. For an $n$-qubit operator, if it is $k$-local, then the number of non-zero Pauli terms is at most $N(n,k,3)=\sum_{l=0}^k C^l_n 3^l\leq \frac{n^k}{2k!}(3^{k+1}-1)$ which is polynomial. We can set $L^\dag L$ to be $k$-local and $L$ to be $k/2$-local to continue our discussion. For $L^\dag L$, note that since the Hamiltonian $L^\dag L$ has the exchange time reversal symmetry as discussed above, thus, when $L^\dag L$ is expressed under the Pauli basis:
\begin{eqnarray}\label{ba2}
L^\dag L=\sum_{i,j=1}g_{ij} P_i\otimes P_j
\end{eqnarray}
where the symbol $\otimes$ separates the row subsystem and the column subsystem, we must have $g_{ij}=g_{ji}^*$. This means that if $L^\dag L$ is an $n$-qubit $k$-local Hamiltonian, the number of the degree of freedom is $N(n,k,3)/2$. For $L$, we will assume that {$X$, $Y$, $Z$, $\sigma^+$, $\sigma^-$} quantum channels may occur for each qubit, which basically contains the majority of the noise channel such as amplitude damping/amplification, dephasing and depolarizing. Thus, for an $n/2$-qubit $k/2$-local LME:
\begin{equation}\label{ba3}
\frac{d\rho}{dt}=\mathcal{L}[\rho]=-i[H,\rho]+
\sum_i \lambda_i(F_i\rho F_i^\dag-
\frac{1}{2}\{\rho,F_i^\dag F_i\})
\end{equation}
there are $N(n/2,k/2,5)$ degrees of freedom of quantum jumps (considering collecting channels) i.e. the number of $\lambda_i$ is $N(n/2,k/2,5)$ and $N(n/2,k/2,3)$ degrees of freedom for $H=\sum_i h_i P_i$. Our task is to get the values of $h_i$ and $\lambda_i$ such that the corresponding $L$ can generate $L^\dag L$. This corresponds to solving a multivariate quadratic equations where the number of unknowns ($h_i$, $\lambda_i$) is $N(n/2,k/2,5)+N(n/2,k/2,3)$ and the number of equations is $N(n,k,3)/2$. However, this is not finished yet since the constraints of $\lambda_i\geq 0$ are not considered. To take these constraints into consideration, we can further introduce another $N(n/2,k/2,5)$ unknowns $w_i$ and $N(n/2,k/2,5)$ equations $w_i^2=\lambda_i$. Thus, in the end, the multivariate quadratic equations has $N_u=2N(n/2,k/2,5)+N(n/2,k/2,3)$ unknowns and $N_e=N(n,k,3)/2+N(n/2,k/2,5)$ equations. In Ref. \cite{courtois2000efficient}, authors introduced a classical algorithm called the XL algorithm to solve the multivariate quadratic equations. The complexity analysis of the XL algorithm states that in the very over-defined regime where the number the unknowns $N_u$ and the number of the equations $N_e$ satisfy $N_e\geq N_u^2 r$ for all $0<r\leq 1/2$, the runtime of the XL algorithm is about $N_u^{\mathcal{O}(1/\sqrt{r})}$. In our case, we find that the equations indeed belongs to the very over-defined regime. Numerical experiments show that for fixed $k$, as $n$ grows, the ratio converges exponentially fast to a fixed value among $(0,1/2]$. We can define the asymptotic ratio between $N_e$ and $N_u^2$ as:
\begin{equation}\label{ba4}
r(k)=\lim_{n \to \infty} \frac{N_e}{N_u^2}=\lim_{n \to \infty}  \frac{N(n,k,3)/2+N(n/2,k/2,5)}{(2N(n/2,k/2,5)+N(n/2,k/2,3))^2}
\end{equation}
Our results show that $r(4)=0.031$, $r(6)=0.015$, etc. which means the runtime of the XL algorithm is about $N_u^{\mathcal{O}(5.678)}$ for 4-local $L^\dag L$ and $N_u^{\mathcal{O}(8.111)}$ for 6-local $L^\dag L$, etc. Since $N_u=2N(n/2,k/2,5)+N(n/2,k/2,3)$ is of order $\mathcal{O}(poly(n))$, we can conclude that under the locality constraints, the runtime of obtaining $L$ from $L^\dag L$ is polynomial.

We further considered a concrete example where the structure of $L:=L(\{\lambda_i\})$ is fixed except for its set of parameters $\{\lambda_i\}$. In this way, $L^{\dagger}L(\{\lambda_i\})$ can be easily obtained and for an arbitrarily chosen $\tilde{L}^{\dagger}\tilde{L}$ of the form in Eq.(\ref{ba2}), we can obtain a set of degree-2 polynomial equations by comparing the terms in $L^{\dagger}L(\{\lambda_i\})$ to terms in $\tilde{L}^{\dagger}\tilde{L}$. We can then use the XL algorithm to solve for the parameters $\{\lambda_i\}$ to see whether there exists a certain set of parameters $\{\lambda_{i0}\}$ such that $\tilde{L}^{\dagger}\tilde{L}=L^{\dagger}L(\{\lambda_{i0}\})$. To be more detailed, $L$ is chosen as
\begin{equation}\label{XXZ}
    \mathcal{L}[\rho]:=-i[H_{\text{XXZ}},\rho]+\sum_{i}\lambda_i (\sigma_{+}^i\rho \sigma_{-}^i-\frac{1}{2}\{\rho,\sigma_{-}^i\sigma_{+}^i\})
\end{equation}
where $H_{\text{XXZ}}=\sum_{\langle ij\rangle}\lambda_z^{\langle ij\rangle}\sigma_z^i\sigma_z^j+\lambda^{\langle ij\rangle}(\sigma_x^i\sigma_x^j+\sigma_y^i\sigma_y^j)$ is the nearest neighbor XXZ model and the upper index of a sigma operator indicates the site it acts on. We further restrict to an one dimensional system with open boundary condition. To conveniently illustrate that the XL algorithm can actually provide correct answers and to study whether it is efficient, we  assign the parameters $\{\lambda_z^{\langle ij\rangle}\}\cup \{\lambda^{\langle ij\rangle}\}\cup \{\lambda_i\}$ with random values and obtain the explicit expression of $L^{\dagger}L$ and also a set of degree-2 equations by comparing $L^{\dagger}L$ with $L^{\dagger}L(\{\lambda_z^{\langle ij\rangle}\}\cup \{\lambda^{\langle ij\rangle}\}\cup \{\lambda_i\})$. In this way, we may solve for parameters and check whether the result is the same as the values we assigned initially.

\begin{figure}[htb]
    \centering
    \subfigure[]{\includegraphics[width=0.48\textwidth]{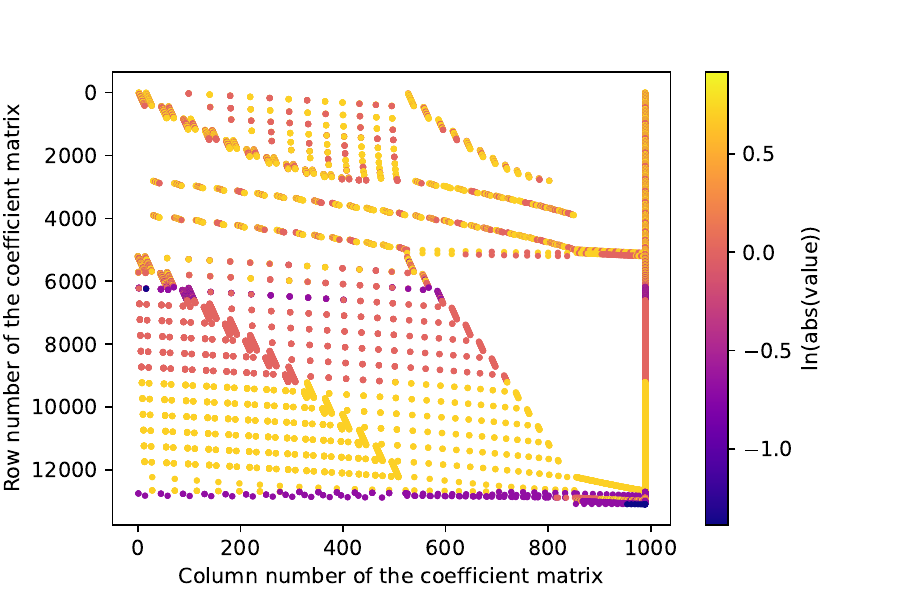}}\label{fig:Matrix}
    \hfill
    \subfigure[]{\includegraphics[width=0.48\textwidth]{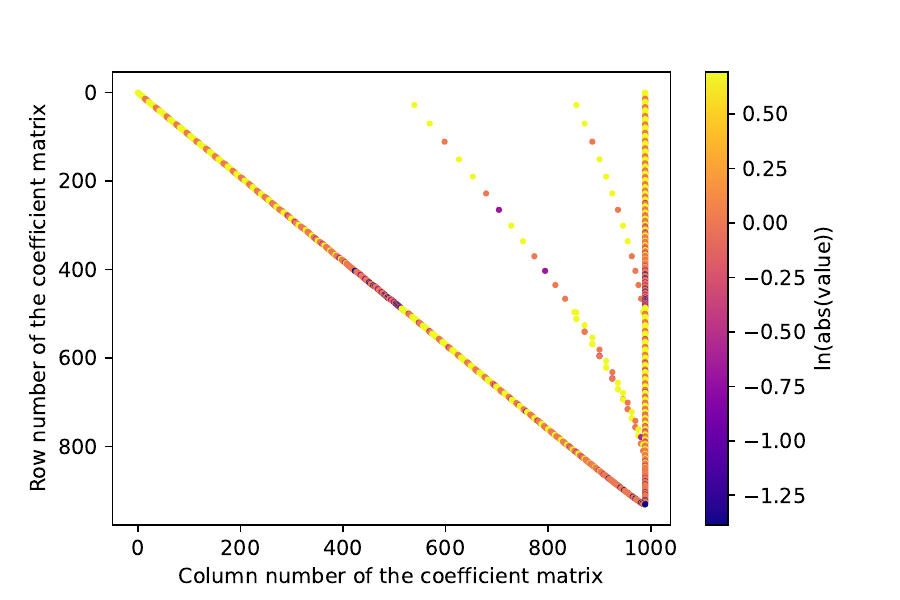}}
    \caption{Coefficient matrix of a set of linear equations. Nonzero values in the matrix are plotted as dots. The color of each dot is determined by the logarithm of its corresponding absolute value. (a) The coefficient matrix of $\mathcal{I}'$ after linearization where $\mathcal{I}'$ is the extension of $\mathcal{I}$ with $D=2$ and $\mathcal{I}$ corresponds to a particular XXZ model under decay described in Eq.(\ref{XXZ}) with 15 lattice sites. (b) The coefficient matrix of the same linearized $\mathcal{I}'$ after a Gaussian elimination step.}  
    \label{fig:Coeff_mat}
\end{figure}
\begin{figure}[!htb]
    \centering
    \includegraphics[width=0.6\textwidth]{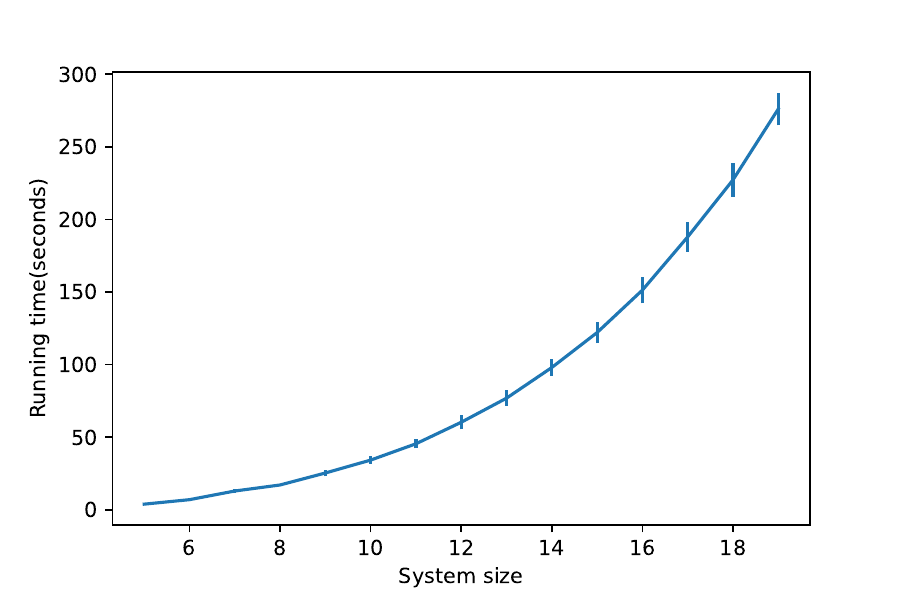}
    \caption{Running time of the XL algorithm at different system sizes. The system size refers to the number of spins in the model described in Eq.(\ref{XXZ}).  Error bars are plotted. }
    \label{fig:Running time}
\end{figure}
Now, before displaying the final results, we discuss briefly the XL algorithm. Consider a set of degree-2 multivariate polynomial equations $\mathcal{I}:=\{f_l(x_1,\cdots,x_n)=0|l\in\Lambda, \deg(f_l)\leq 2\}$ where $\Lambda$ is a finite set. Directly solving $\mathcal{I}$ is usually hard. However, if we were able to obtain from $\mathcal{I}$ some univariate polynomial equations, we would be able to solve for one variable easily and reduce the number of variables in $\mathcal{I}$ by 1. If the above process can be repeated, we would eventually solve $\mathcal{I}$. The XL algorithm provides us with an approach to implement the above procedure. The XL algorithm consists of the following steps
\begin{itemize}
    \item Step 1.(Extension) Fix an integer $D\geq 2$. Extend $\mathcal{I}:=\{f_l(x_1,\cdots,x_n)=0|l\in\Lambda, \deg(f_l)\leq 2\}$ to a new set $\mathcal{I}'$ such that $\mathcal{I}'=\{f_l(x_1,\cdots,x_n)\cdot g(x_1,\cdots,x_n)=0|f_l\in\mathcal{I}, g\in\mathrm{Mono}(D-2)\}$ where $\mathrm{Mono}(D-2)$ is the set of all monomials of $(x_1,\cdots,x_n)$ with degree less or equal than $(D-2)$. In this way, $\forall$ element in $\mathcal{I}'$ has degree $\leq D$. And, obviously, if we can solve $\mathcal{I}'$, we would easily obtain the solution of $\mathcal{I}$. 
    \item Step 2.(Linearization) Now, treat each monomial in $\mathrm{Mono}(D)$ as an independent variable. In this way, we can treat $\mathcal{I}'$ as a set of linear equations. 
    \item Step 3.(Gaussian elimination) Using Gaussian elimination, we may obtain a set of equations that only contains monomials of a single variable. If this step failed, return to step and choose a larger $D$, otherwise, we can easily solve for this single variable. Then, we can simplify $\mathcal{I}$ and $\mathcal{I}'$ by inserting the value of the solved variable. Usually, no further extension is required, and we may simply proceed to step 2 to obtain a new set of univariate polynomial equations.
\end{itemize}

We want to mention that since we are solving a class of Hamiltonians with known ground energy, our result doesn't violate that finding k-local Hamiltonian's ground energy is QMA-hard \cite{kempe2006complexity,gharibian2014quantum}. We want to generalize this point that the ground energy $E_0$ is not necessarily needed for the XL algorithm part since we can just abandon the equation corresponding to the identity term. If we do so, and the XL algorithm gives a solution, then this means that the Hamiltonian can be solved and the ground energy can be calculated from the solution. However, this also means that a Hamiltonian that we can solve must have its ground energy classically easy to get, which we can still see as the known energy case.

As an illustration of the XL algorithm, we specifically choose a 15-qubit 1D XXZ model under decay described in Eq.(\ref{XXZ}) and obtain a set of polynomial equations $\mathcal{I}$ by assigning all parameters as 1 and 
comparing the terms in $L^{\dagger}L(\{\lambda_z^{\langle ij\rangle}=1\}\cup \{\lambda^{\langle ij\rangle}=1\}\cup \{\lambda_i=1\})$ with $L^{\dagger}L(\{\lambda_z^{\langle ij\rangle}\}\cup \{\lambda^{\langle ij\rangle}\}\cup \{\lambda_i\})$. Then, following the XL algorithm, we obtain $\mathcal{I}'$ by extending $\mathcal{I}$ with $D=2$ and then perform the linearization step on $\mathcal{I}'$. The resulting coefficient matrix of the set of linear equations in linearized  $\mathcal{I}'$ is shown in Fig.(\ref{fig:Coeff_mat}a). We then perform Gaussian elimination on the linearized $\mathcal{I}'$ and the resulting coefficient matrix is shown in Fig.(\ref{fig:Coeff_mat}b). According to Fig.(\ref{fig:Coeff_mat}b), we have obtained univariate polynomial equations after the Gaussian elimination procedure. By comparing the number of rows before and after Gaussian elimination, we can see that there are a lot of redundant equations in $\mathcal{I}'$. Also, from Fig.(\ref{fig:Coeff_mat}a), we can see that the coefficient matrix is sparse so that the Gaussian elimination procedure can be carried out efficiently. 

To exhibit the performance and scaling property of the XL algorithm, we gradually increase the lattice size $N$ from $5$ to $19$ and for each $N$, we repeatedly solve for 25 times $\mathcal{I}$ obtained from $L^{\dagger}L$ whose parameters are sampled independently according to uniform distribution on $(0,1)$ and record the running time for each iteration. See Fig.(\ref{fig:Running time}) for the result.  

The program can be found online \cite{XLLdagL2022}.
\end{document}